# Characterizing virulence differences in a parasitoid wasp through comparative transcriptomic and proteomic

Samuel Gornard[1], Pascaline Venon[1], Florian Lasfont[1], Thierry Balliau[2], Laure Kaiser[1], Florence Mougel[1,*]

[1] EGCE, Université Paris-Saclay, CNRS, IRD, UMR Évolution, Génomes, Comportement et Écologie, 91190 Gif-sur-Yvette, France

[2] PAPPSO, Université Paris-Saclay, INRAE, CNRS, AgroParisTech, GQE - Le Moulon, 91190, Gif-sur-Yvette, France

* Corresponding author. E-mail address: florence.mougel-imbert@universite-paris-saclay.fr (F. Mougel).

**Abstract**

**Background**: Two strains of the endoparasitoid *Cotesia typhae* present a differential parasitism success on the host, *Sesamia nonagrioides*. One is virulent on both permissive and resistant host populations, and the other only on the permissive host. This interaction provides a very interesting frame for studying virulence factors. Here, we used a combination of comparative transcriptomic and proteomic analyses to unravel the molecular basis underlying virulence differences between the strains.

**Results**: First, we report that virulence genes are mostly expressed during the nymphal stage of the parasitoid. Especially, proviral genes are broadly up-regulated at this stage, while their expression is only expected in the host. Parasitoid gene expression in the host increases with time, indicating the production of more virulence factors. Secondly, comparison between strains reveals differences in venom composition, with 12 proteins showing differential abundance. Proviral expression in the host displays a strong temporal variability, along with differential patterns between strains. Notably, a subset of proviral genes including protein-tyrosine phosphatases is specifically over-expressed in the resistant host parasitized by the less virulent strain, 24 hours after parasitism. This result particularly hints at host modulation of proviral expression.

**Conclusions**: This study sheds light on the temporal expression of virulence factors of *Cotesia typhae*, both in the host and in the parasitoid. It also identifies potential molecular candidates driving differences in parasitism success between two strains. Together, those findings provide a path for further exploration of virulence mechanisms in parasitoid wasps, and offer insights into host-parasitoid coevolution.

# Background

Endoparasitoid wasps are hymenopterous insects that lay their eggs inside the hemocoel of their host (Pennacchio & Strand, 2006). To allow full development of the embryos, the mother wasp must avoid the immune reaction of the host, which consists of encapsulation of the parasitoid eggs or larvae (Lavine & Strand, 2002). Immunity bypassing is achieved either by passive evasion (Hu et al., 2003; Quicke, 2014) and/or by the injection of maternal substances in the host alongside the eggs. Those factors participate in the virulence of the wasp, i.e. its ability to successfully parasitize its host. They can act alone or in synergy in order to manipulate host physiology and alter immunity (Asgari, 2006; Mabiala-Moundoungou et al., 2010). Like most hymenopterans (Walker et al., 2018), endoparasitoid wasps synthesize venom in a dedicated gland and the fluid is stored in the venom sac and is injected during oviposition. Venom starts to be produced at the pupal stage (Jones & Wozniak, 1991) and is mostly made of proteins of various sizes (Leluk et al., 1989; Moreau & Asgari, 2015) with frequent glycosylation. Small peptides can be found, even if less studied (Hauser et al., 2010). Venom components encompass a great variety of functions (Rivers, 2007) and likely play an important role in immunosuppression after injection in the host (Er et al., 2011). Venom can prevent melanization (Asgari et al., 2003) and encapsulation by reducing the number of circulating hemocytes (Uçkan et al., 2010), disrupting their spreading abilities (Furihata et al., 2013; Yu et al., 2007) or even inducing apoptosis (Rivers et al., 2009). It can also induce metabolic changes to allow for a better feeding of the parasitoid's offspring (Danneels et al., 2010; Rivers & Denlinger, 1995; Walker et al., 2018).

Several lineages of parasitoid wasps also harbour domesticated viruses in their genomes. Wasps from the Ichneumonoidea superfamily are well known for the polyDNAviruses that are found within several distinct subfamilies, hinting towards convergent acquisition of the virus. Ichnovirus are present in Campopleginae and Banchinae Ichneumonidae subfamilies, and bracovirus in the six subfamilies of the Braconidae Microgastroid complex (Lapointe et al., 2007; Whitfield & O'Connor, 2012). The polyDNAviruses exist as proviruses inside the genome of the wasp, but replicate only in cells of the female calyx wall (Beckage & Drezen, 2012). After replication, proviral circles are packaged in virions produced in the same cells by integrated genes originating from nudiviruses or baculoviruses (Strand & Burke, 2020). Viral particles are part of what is called the calyx fluid, in which they can cover eggs with a protective coating allowing passive evasion reported in some species (Asgari & Schmidt, 1994; Quicke, 2014). After injection alongside the eggs, they infect host cells which in turn will produce virulence factors (Moreau et al., 2009). This will first manipulate host development (Soller & Lanzrein, 1996) through endocrine regulation (Beckage, 2012) in order to allow complete development of the wasp's offspring (Whitfield et al., 2018). Virus will also severely impair immunity by disrupting encapsulation of the wasp eggs and larvae, and capsule melanization (Strand, 2012b). Together with other virulence factors, such as Virus-Like Particles or teratocytes (Quicke, 2014), venom and polyDNAviruses are key factors for successful parasitism.

*Cotesia typhae* is an eastern African Braconidae endoparasitoid wasp belonging to the Microgastroid complex. This gregarious species is strictly specialized on the larvae of the lepidopteran stem borer *Sesamia nonagrioides* (Kaiser et al., 2017), whose African and European populations diverged 180,000 years ago and accumulated genetic differentiation (Kergoat et al., 2015; Moyal et al., 2011). Two populations of *C. typhae* were sampled in Kenya, west and east of the oriental rift valley, respectively in Kobodo and Makindu localities. Laboratory strains were initiated from samples of these two localities, and they differed in their ability to successfully parasitize the French host population (Benoist et al., 2017). While both scored relatively good parasitism success on the susceptible Kenyan population (termed SnR-, 88 and 94%, respectively), the Makindu strain appeared to be less virulent on the resistant French population (termed SnR+), reaching 30% success, against 94% for the Kobodo strain. As the Makindu strain appeared to be less virulent, it was named CtV-, and by contrapositive, Kobodo strain was named CtV+. They were both reared as iso-female strains for genetic studies, at the ICIPE (Kenya) and EGCE (France) laboratories, where they conserved their virulence differences (Benoist, Paquet, et al., 2020; Gornard, unpublished data, 2021). Wide genomic regions were shown to be involved in the difference of virulence, and they contain both venom synthesis, proviral and nudiviral genes (Benoist, Capdevielle-Dulac, et al., 2020). Moreover, the expression of two proviral genes, CrV1 and Cystatin, differ between the strains. Parasitism by CtV+ females leads to higher expression levels of these two genes in the French host (Benoist et al., 2017). Thus, virulence differences could potentially be linked to venom synthesis and/or proviral gene expression. Venoms are known to be able to evolve rapidly in adaptation to specific hosts (Cavigliasso et al., 2019;

Colinet et al., 2013; Mathé-Hubert et al., 2019), even between close strains of the same species. Evolution in protein composition or relative abundance is therefore often a sign of host adaptation rather than phylogenetic differentiation (Poirié et al., 2014), and rapid changes can be achieved by simple gene expression regulation, inducing quantitative differences (Cavigliasso et al., 2019). In polyDNAvirus-bearing parasitoid species, venom can simply be devoid of function, but it can also overlap and even synergise with the effect of the virus (Moreau & Asgari, 2015). More specifically, in the genus *Cotesia*, venom appears to be crucial for entry of the virus in host cells and expression of virulence genes (Asgari, 2006; Zhang et al., 2004). Thus, venom may not be able to suppress the immune reaction alone, but can modulate the effect of the virus to achieve it.

Given the interaction between these two virulence factors, the virulence of *C. typhae* and the differences between CtV+ and CtV- strains was studied through venom composition and gene expression. Recent experiments showed that the encapsulation of CtV- eggs by the French host population started after 24h post-oviposition and was completed at 96h (Gornard et al., 2024), a time window appropriate to investigate more precisely gene expression in the host. By identifying candidate proteins and genes responsible for this virulence difference, we will better understand how host specialization can occur in relation with venom compounds and viral expression.

So this study aims to characterize the molecular basis of virulence, the dynamic of virulence factors expression and the virulence differences between the two parasitoid strains by combining transcriptomic and proteomic approaches (Fig. 1).

# Methods

1. Insect rearing

The parasitoid wasp *Cotesia typhae* and its host *Sesamia nonagrioides* were reared as described in Gornard et al. (2024). Briefly, *S. nonagrioides* larvae were kept on an artificial diet until pupation, at 26°C, 70% relative humidity, and a 16/8 light/dark cycle. Chrysalids were collected and placed in boxes where adults could emerge, mate, and lay eggs (21°C, 70% RH, 16/8 light cycle). Two populations were reared separately: a Kenyan one (less resistant, SnR-), originating from individuals collected in Makindu and Kabaa (south-east of Kenya), and a French one (more resistant, SnR+), from individuals collected in the south-east of France and refreshed yearly with wild larvae.

One to three-day-old *C. typhae* adults (70% RH, 12/12 light/dark cycle) were used to parasitize L5 SnR- larvae previously fed with fresh maize. Parasitized larvae were fed with the artificial diet for 12 days (26°C, 70% RH, 16/8 light/dark cycle), and then placed on tissue paper to allow parasitoid egression. Cocoon masses were then retrieved and kept in plastic boxes for the next generation. Two diverging inbred strains, CtV+ (more virulent, Kobodo) and CtV- (less virulent, Makindu), were reared separately and named after the Kenyan locality where they were first collected.

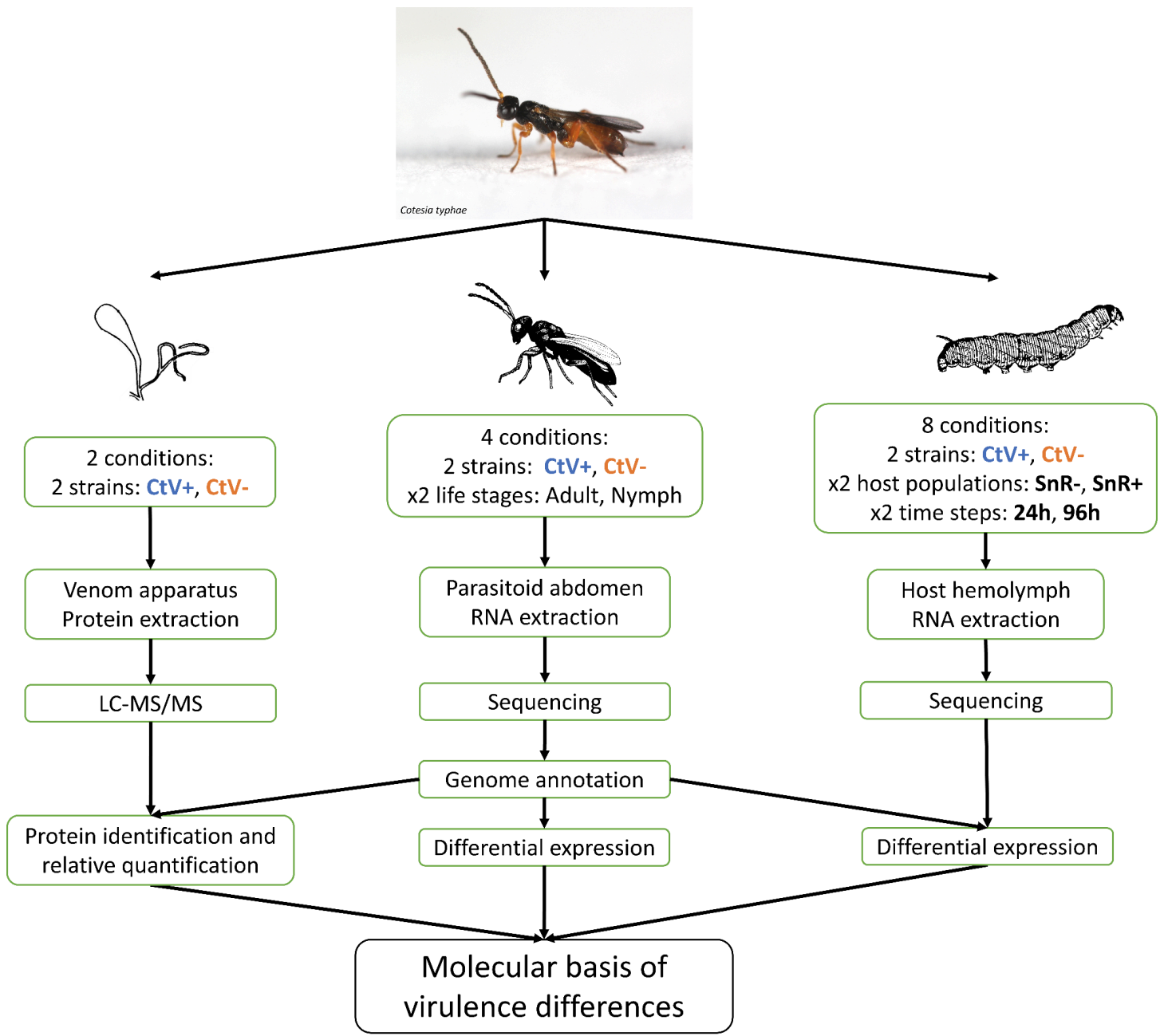

Figure 1: Diagram of strategies used for these experiments. Proteomic identification of venom content was confronted with RNA sequencing of both host hemolymph and parasitoid in order to identify candidate genes in virulence difference. Inspired by Zhao et al., (2023).
Copyrights: Paul-André Calatayud, Greany et al. (1984), canva.com, Rincon-Vitova Insectaries.

## 2. RNA extraction

### 2.1. Free stages

Total RNA was extracted from the abdomen of wasps of both CtV- and CtV+ strains using the nucleospin RNA set for Nucelozol® (Macherey-Nagel) and following the manufacturer's instructions. Nymphs 24 hours before emergence and adults on the day of emergence were used to account for life-stage gene expression variation. RNA extraction was performed on a pool of 20 female abdomens for each biological replicate, and each pool consisted of individuals coming from four different cocoon masses. Three biological replicates per condition were made, leading to a total of 12 samples. Dosage and quality control were performed by Nanodrop™ (ThermoScientific), RNA Broad Range Qubit® kit (Invitrogen), and automated electrophoresis with Experion RNA StdSens analysis chip (Bio-Rad). All samples were stored at -80°C before being sent to the Novogene company for cDNA library construction and transcriptome sequencing (Illumina technology, paired-end, length of 150 base pairs, and sequencing depth of 40X). This dataset will be referred to as free-stage experiment (FSE).

### 2.2. Parasitic stages

Total RNA was extracted from the hemolymph of L5 SNK and SNF parasitized by both strains of *C. typhae*, either 24 or 96 hours after parasitism, corresponding respectively to embryonic and first larval stage, using the same kit as described before. Host larvae were killed by exposure to -80°C for five minutes, and then surface sterilized with 96° ethanol. Hemolymph was collected by bleeding from a proleg and stored at -80°C until RNA extraction. Each larva was dissected after bleeding to verify the presence of *C. typhae* eggs or larvae, and thus ensure that parasitism took place (see Gornard et al., 2024, for details). RNA extraction was performed on 60 µL pools made of 10 µL of hemolymph from 6 larvae. Three biological replicates per condition were made, leading to a total of 24 samples. Dosage and quality control were performed by RNA Broad Range Qubit® kit (Invitrogen), and automated electrophoresis with Experion RNA StdSens analysis chip (Bio-Rad). Samples were stored and sequenced as described before. This dataset will be referred to as parasitic-stage experiment (PSE).

## 3. Genome re-annotation and functional annotation

Raw reads from the free stages were cleaned using Cutadapt v. 1.15 (Martin, 2011) in order to remove PCR adapters, low-quality bases (q < 28), unidentified bases (labeled as N), and reads shorter than 40 base pairs. Cleaned reads were mapped to *C. typhae* genome (Muller, Chebbi, et al., 2021) using STAR v. 2.5.3a (Dobin et al., 2013). Reads that mapped to more than five different loci and mappings with more than three mismatches were eliminated. Introns size was allowed between 10 and 50,000 base pairs long and maximum distance between two paired reads was fixed to 50,000 base pairs. Then, the consensus transcriptome was obtained with Cufflinks v. 2.2.1 (Ghosh & Chan, 2016; Trapnell et al., 2012), by building the transcriptome of each of the 12 samples before merging them with Cuffmerge. *C. typhae* genome was then re-annotated with MAKER v. 2.31.10 (Campbell et al., 2014), following the protocol detailed in Muller, Ogereau, et al. (2021). Briefly, after identifying repeated elements with RepeatModeler v. 2.0.1 (https://github.com/Dfam-consortium/RepeatModeler), we used *C. typhae* consensus abdominal transcriptome to guide the first round of MAKER, along with proteomes of five hymenopteran species: *Chelonus insularis* (Hymenoptera: Braconidae) (NCBI: GCF_013357705.1), *Cotesia glomerata* (Hymenoptera: Braconidae) (NCBI: GCF_020080835.1), *Microplitis demolitor* (Hymenoptera: Braconidae) (NCBI: GCF_000572035.2), *Nasonia vitripennis* (Hymenoptera: Pteromalidae) (Uniprot ID: UP000002358; Dalla Benetta et al., 2020) and *Apis mellifera* (Hymenoptera: Apidae) (NCBI: GCF_003254395.2). Selected proteomes had a complete BUSCO of more than 95% (BUSCO v. 5.0.0; Manni et al., 2021) with the hymenopteran gene set. Three more rounds of MAKER were conducted with two gene predictors, SNAP and Augustus. Finally, newly found genes and their associated proteins were automatically annotated by blastp (v. 2.10.1) search against the non-redundant database UniProtKB/Swiss-Prot database and the proteome of *Cotesia congregata* (Uniprot ID: UP000786811). Conserved domains were annotated with InterProScan v. 5.46-81.0. The quality of the annotation was evaluated with the AED (Annotation Edit Distance) that was calculated for new gene models (Campbell et al., 2014; Holt & Yandell, 2011). Models obtained through the passing of former ones from an ancient annotation were assigned an AED value of 1.

Further characterization of the proteome was performed with three methods. First, GO terms and KEGG onthology were identified with eggNOG-mapper v. 2.1.9 (Cantalapiedra et al., s. d.), using the Diamond aligner and the default settings (thresholds: e-value = 0.001, score = 60, and identity percentage = 40). Second, putative venom proteins were identified with blastp search against the manually curated hymenopteran venom database iVenomDB (Chen et al., 2023), which contains 4,847 proteins from 139 insect species. We set the e-value threshold to $10^{-6}$ and only kept the highest match. Last, excretion signal peptides were identified with Signal-P v. 5.0b (Almagro Armenteros et al., 2019).

We then crossed the annotation with the localization of the proviral segments (identified by Muller, Chebbi, et al., 2021) and the QTL involved in parasitism success (identified by Benoist et al., 2020) to catalog potent virulence genes. Novel nudiviral genes were annotated through blastp with *C. congregata* genome.

### 4. RNA-seq data analysis

Raw reads of both experiments were cleaned using Cutadapt v. 1.15 and 1.18 with the same parameters as above (see 2.3). Cleaned reads of each experiment were mapped on the re-annotated *Cotesia typhae* genome using STAR v. 2.5.3a and the same parameters as above (see 2.3). The number of reads mapping on each gene was obtained with the option -quantMode geneCounts. Reads from the free-stage experiment that mapped on the proviral segments identified by Muller, Chebbi, and collaborators (2021) were screened with IGV v. 2.16.1 (Robinson et al., 2011) in order to check for genomic contamination.

Expression matrices of both experiments were filtered for low counts by HTSFilter R package v. 1.34.0 (Rau et al., 2013), and differential expression analysis was made with DESeq2 R package v. 1.34 (Love et al., 2014). Because of the consequent difference in *C. typhae* gene expression between 24 and 96 hours, those two conditions were mostly analyzed separately and referred to as 24-hour parasitic-stage experiment (24-PSE) and 96-hour parasitic-stage experiment (96-PSE). Samples of the FSE were analyzed with a model of type expression ~ strain*stage, and samples of the PSE with a model of type expression ~ strain*host*time, before being split into two datasets analyzed separately with models of type expression ~ strain*host.
Sample expression profiles were compared using the PCA function and VST (Variance Stabilizing Transformation) normalization of DESeq2 to control for outliers and remove abnormal samples (CtV-/SnR+/96h/3 was consequently removed from subsequent analyses). Outside of the time condition comparison, expression matrices of the 24-PSE and 96-PSE were analyzed separately with DESeq2 to eliminate the bias induced by the very low number of *C. typhae* genes expressed at 24 hours (54 genes expressed after filtering).
Genes were considered as differentially expressed between conditions when their adjusted p-value was lower than 0.05.

The expression profile of some relevant subsets of genes was analyzed by hierarchical clustering, using the pheatmap R package (Kolde, 2019) with the manhattan clustering distance method. Expression levels were scaled by row in order to account for between-gene expression variation.

### 5. Venom gland collection

Venom was extracted from venom gland and reservoir (venom apparatus) of two to three-day-old female *C. typhae* from both strains. Six replicates of 15 venom apparatuses per strain were sampled. Females were anesthetized on ice and their abdomen was torn open with dissection forceps in a drop of sterile phosphate buffer solution (PBS) 1X. Venom apparatuses were then isolated and gathered in 15 µL of sterile PBS 1X, before being dilacerated with a needle and the tip of a forceps. Venom preparation was centrifugated at 12,000 G, 4°C for 4 minutes and only supernatant was collected to discard cellular residues. All 12 samples were stored at -80°C until further analysis.

### 6. Proteomic analysis

In order to visualize qualitative differences between CtV+ and CtV- strains, one sample of solubilized venom extract per strain was separated with SDS-PAGE in NuPAGE® MOPS-PBS buffer, stained with Coomassie and compared to SeeBlue™ Plus2 (ThermoScientific) protein ladder.

The ten samples left were used for LC-MS/MS proteomic analysis. Briefly, samples were solubilized in Laemmli buffer (2 % SDS, 10 % Glycerol, 5% betamercaptoethanol, 0.002 % bromophenol blue, 62.5mM Tris-HCL pH 6.8) and 10 µL of each sample (corresponding to 10-females equivalent) were deposited on electrophoresis gel. After short migrations, bands were cut and separated into two fractions depending on molecular weight (light and heavy). Proteins were reduced with DTT, alkylated with iodoacetamide, and digested with trypsin in gel as described in (Recorbet et al., 2021). Samples were analyzed on a TimsTof Pro mass spectrometer (Bruker, Billerika USA) coupled to a nanoElute chromatography (Bruker). They were then loaded on a nanoEase trap column and eluted on an aurora column. Analysis details are given in supplementary material 3.

Identification was performed using X!Tandem software (Craig et al., 2004; version 2015.04.01.1) against the *Cotesia typhae* database from NCBI (txid 2053667 , 8423 entries), a custom contaminant database (keratins, trypsin, ...) and the newly annotated database (18667 entries) with parameters detailed in supplementary material 3.

Protein inference was performed using i2masschroq software (Langella et al., 2017; version 0.4.76). A protein was validated if it was identified by at least two distinct peptides with an e-value smaller than 0.01 and a resulting e-value for the protein smaller than 0.00001 using all samples together. Using *in silico* generated decoy database, FDR was estimated to 0.23 % for peptide spectrum match and 0.36 % for protein identification.

Extracted Ion Current quantification was performed using masschroq (Valot et al., 2011; version 2.4.20,). Mass precision was set to 20 ppm and quantification was performed on 80 % of the theoretical natural isotopic profile.

Quantitative differences between strains were analyzed with the R package MCQR v. 0.6.9 (PAPPSO©, https://forgemia.inra.fr/pappso/mcqr). First, the two fractions were filtered separately for peptides with too much variation in their retention time (cutoff = 17 seconds), before being fused together. The analysis was carried out on proteins that were either identified as putative venom by blast against the iVenomDB or that likely bore an excretion signal peptide. Data was normalized using the median of differences method by using one of the bulk, after which all bulks were removed for subsequent analysis. Peptides that were shared by two or more sub-groups of proteins, peptides-mz for which the proportion of missing values exceeded 20%, and those which intensity profiles did not correlate with those of any others belonging to the same protein were removed from the dataset. Protein abundance was calculated as the sum of peptide intensities, after imputation of missing values. Missing abundance protein values were replaced by the minimum obtained for each protein in the whole experiment, and iBAQ index (Fabre et al., 2014) was calculated to compare protein abundances. Proteins showing a fold change between the two conditions inferior to 1.5 were discarded. Then, we used an ANOVA to compare protein quantification between strains. Proteins with adjusted p-value < 0.05 were considered as differentially abundant between the two strains.

Differentially abundant proteins were re-blasted against the nr NCBI database.

### 7. Clustering and enrichment analysis

In order to have an insight on the functions of genes and proteins identified in the preceding analyses, we performed gene clustering and enrichment analysis of Gene Ontology terms.

Expression matrices of both 96-PSE and FSE were concatenated in order to cluster the genes based on their expression profiles in several conditions. The clustering was performed by the R package HTSCluster v. 2.0.11 (Rau et al., 2015), using the DESeq normalization, and after filtering with HTSFilter. The 24-PSE was excluded from the clustering analysis due to the bias induced by the very low number of genes expressed in this condition (54 genes expressed after filtering). Genes with a probability of belonging inferior to 0.95 were excluded from their cluster for subsequent analysis.

We also manually selected three subsets of genes based on their biological relevance: venom genes (selected for proteomic analysis), proviral genes and genes over-expressed either in CtV+ or in CtV- in the FSE and the 96-PSE (adjusted p-value ≤ 0.05 and |Log2Fold| ≥ 1).

We performed a GO enrichment analysis on automatically generated clusters and selected subsets of genes. Three separate methods were used to perform Fisher's exact test and only GO terms with adjusted p-values of less than 0.05 in all three methods were considered significantly enriched for said cluster. We used ClusterProfiler v. 4.2.2 (Wu et al.,

2021) with weight01 algorithm, the find_enrichment function of GOATools v. 0.7.11 (Klopfenstein et al., 2018), and the R package TopGO (Alexa & Rahnenfuhrer, 2017).

# Results

## 1. Genome re-annotation

The automatic annotation of *C. typhae* genome performed by MAKER yielded 17,235 genes with a total of 80,084 exons, while the original annotation contained 8,591 genes. The value of complete proteomic BUSCO also improved, reaching 79.6% (S: 70.0%, D: 9.6%, F: 6.0%, M: 14.4%), compared to that of the former annotation (S: 62.8%, D: 0.5%, F: 5.8%, M: 30.9%), on the hymenopteran gene set.

We used the AED metric to assess the quality of the annotation and found that 61.6% of the gene models (including the ones transferred from former annotation) scored an AED of 0.5 or less. We also found that 40.17% of the predicted genes contained a Pfam domain and 47.23% contained an IPR domain.

EggNOG-mapper was able to attribute GO terms to 6,304 proteins and KEGG orthologies to 5,278 proteins (Supplementary material 2). Blasting against the iVenomDB yielded 2,269 potent venom proteins, 5 of which are located in the proviral segments, and 136 in the QTL (Supplementary material 1, Tab 1). Signal-P identified 1,216 proteins that likely bear an excretion signal peptide.

By crossing the annotation with the coordinates of proviral segments, we identified 161 genes located in those segments. Protein blast against the genome of *C. congregata* also revealed 9 genes similar to bracoviral genes, but located outside of those segments. We also crossed the annotation with the QTL involved in parasitism and reproductive success and found 1,448 genes in these loci.

## 2. Transcriptomic data quality analysis

Transcriptome sequencing was performed on 24 and 12 samples corresponding to parasitized host hemolymph and parasitoid abdomen respectively.

Host hemolymph samples contained between 79,898,342 and 135,421,996 raw reads, including both wasp and host RNA transcripts. Reads had a Q30 ranging from 91.58 to 93.26% and a GC content ranging from 39.20 to 45.64%. Trimming and cleaning conserved 99.04% of reads on average. For the 24-hour condition, between 1.50 and 6.26% of the reads mapped on annotated genes, as most of the rest of the reads mapped on the genome of the host due to early parasitism stage. For the 96-hour condition, this value ranged from 25.77 to 52.70%, with the exception of replica CtV-/SnR+/96/3, which scored only 1.14% of mapped reads.

Parasitoid abdomen samples contained between 61,905,750 and 88,871,878 raw reads, with a Q30 ranging from 90.16 to 95.06% and a GC content ranging from 37.72 to 39.23%. Trimming and cleaning conserved 99.60% of reads on average. Between 57.62 and 60.59% of the reads mapped on annotated genes, and between 11.06 and 13.81% mapped outside of annotated loci.

Reads from the FSE that mapped on proviral segments were visualized with IGV in order to verify the presence of intron gaps. The presence of reads cut in half and bordering introns present in proviral genes show that they were obtained from mature mRNA and not through sequencing of genomic contamination. We chose four random proviral genes with strong expression in the FSE and revealed the presence of such intron gaps with the help of sashimi plots (data not shown). Thus, we concluded that proviral gene expression was not caused by genomic contamination.

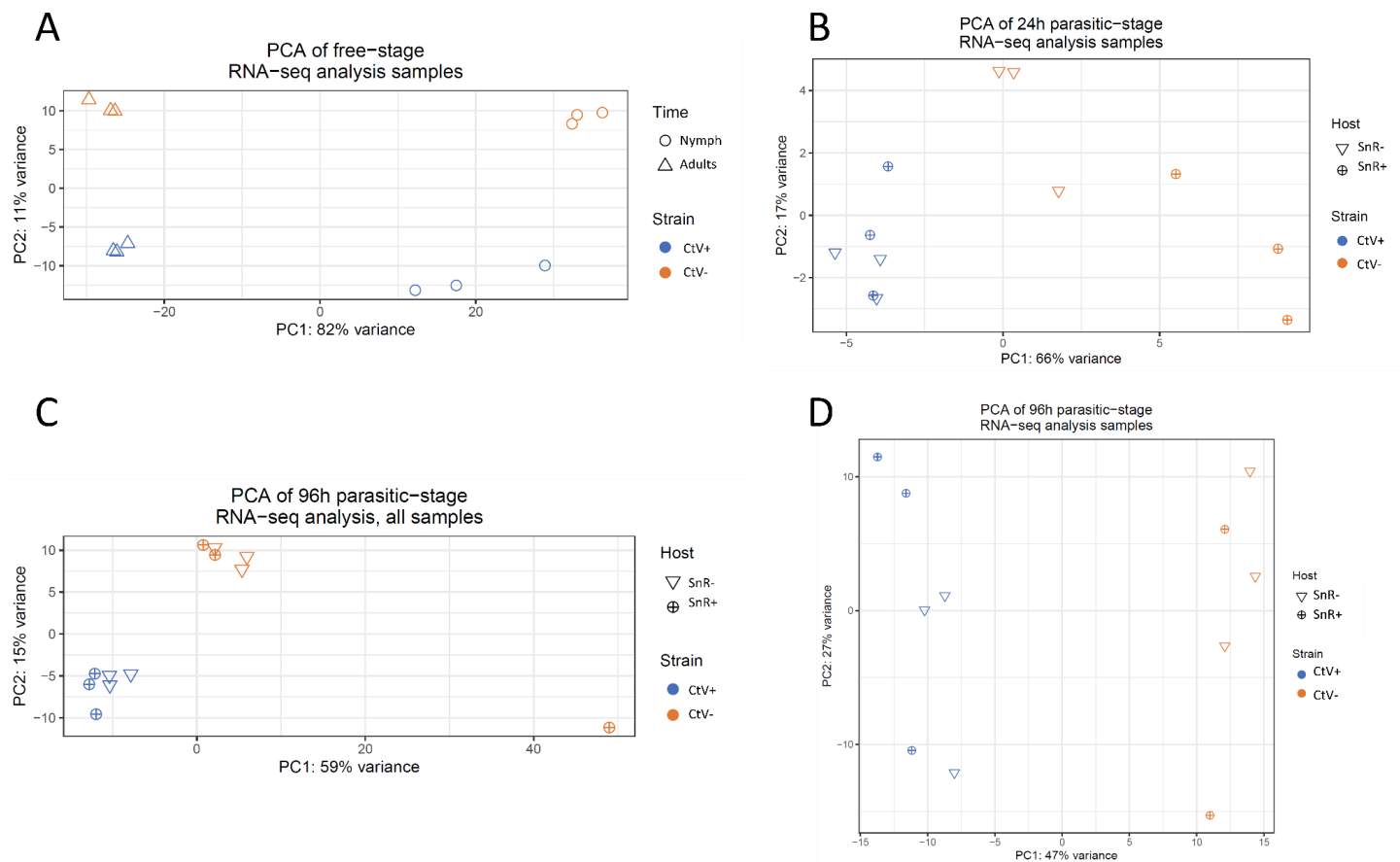

Figure 2: Principal component analysis of gene expression per sample, after normalization by VST. Sample distribution for (A) Free-stage experiment. (B) 24-hour parasitic-stage experiment. (C) All 12 samples of the 96-hour parasitic-stage experiment. (D) 11 samples of the 96-hour parasitic stage experiment.

Gene expression was analyzed with DESeq2 after filtration for low counts by HTSFilter. Out of the 17,235 genes annotated in *C. typhae* genome (see below), HTSFilter conserved 54 genes for the 24-PSE (which were all proviral genes), 5,245 for the 96-PSE, and 8,450 for the FSE.

PCA analysis first confirmed that all replicates of the same condition were closer than samples from different conditions (Fig. 2), with the exception of the replicate CtV-/SnR+/96/3 (Fig. 2, C), which had a very low read mapping percentage on the *C. typhae* gene set. This outlier sample was then excluded from further analysis, and the resulting PCA of 96-PSE gene expression profiles is shown in Fig. 2 (D).

This analysis also gave indications about the main factors driving gene expression profiles. Concerning the FSE (Fig. 2, A), samples were mostly clustered by life stage (adult or nymph) and separated along the first principal component (PC), which represented 82% of the variance. Strain also impacted gene expression profiles, as the second PC (11% of variance) also segregated the samples of the two *C. typhae* strains.

Concerning the 24-PSE (Fig. 2, B), the first PC (66% of variance) separated the samples between strains. The host population did not seem to separate CtV+ samples, but it did so for the CtV- ones, along the second PC (17% of variance). This indicates that expression profiles of CtV+ samples were similar between host populations, but that was not the case for CtV- ones.

Concerning the 96-FSE (Fig. 2, D), strain was again the main factor segregating the samples, along the first PC (47% of variance). Even if the second PC accounted for 27% of the variance, it did not seem to segregate the samples according to the host population, indicating a much broader variation between samples, and no such pattern as in the 24-PSE could be seen.

## 3. Free stages expression
### 3.1. Venom proteomic analysis

Venom samples of both parasitoid strains were separated with SDS-PAGE in order to check for qualitative differences between them. *C. typhae* venom protein ranged from 97 to 14 kDa, with very few proteins outside this range. No major qualitative difference was spotted.

The protein content of both venoms was analyzed with LC-MS/MS. Raw data from MS detection was compared to the putative digested proteomes of *C. typhae* in order to identify proteins. A total of 763 different proteins were identified in the venom of both strains. Out of 763 identified proteins, 601 were present in the newly annotated *C. typhae* proteome, and 162 were found only in the former proteome, showing that the annotation missed some genes. Thus, this new biological data will help us refine the annotation in the future.

Comparative quantitative analysis between strains was performed with MCQR. Filtration for peptides with too much variation in their retention time removed 6 proteins. In order to reduce tissue protein contamination in the samples, the subsequent analysis was restricted to proteins that either were identified as potential venom (iVenomDB blast) or bore an excretion signal. This method selected 362 proteins out of 763, which we referred to as "putative venom". Among them, 276 were newly annotated. Quality control and normalization selected 257 proteins. Abundance was computed as a sum of peptide intensities, which correspond to the area under the spectrography curve and is therefore relative (and has no unit). Between-protein abundances were compared using iBAQ index. The ANOVA yielded 12 proteins that were differentially abundant between the two strains (Fig. 3 and Table 1).

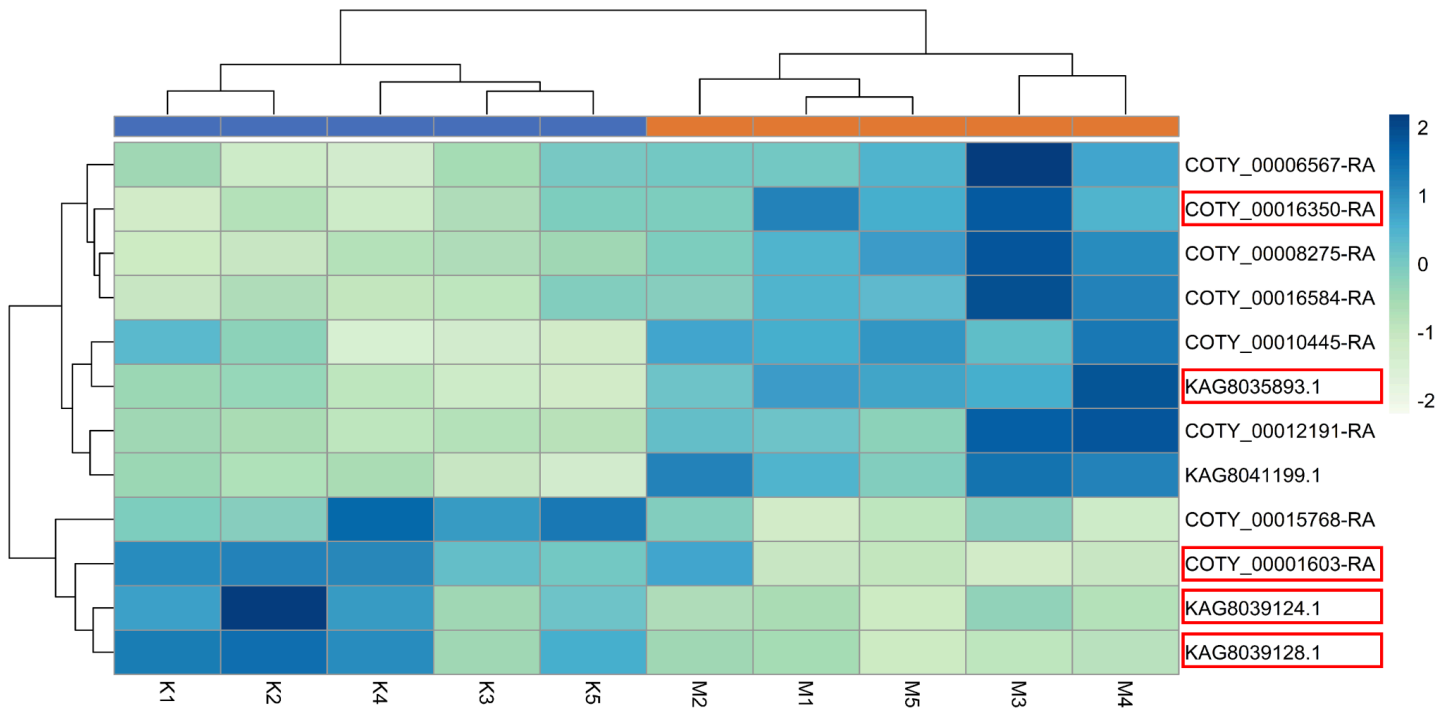

Figure 3: Heatmap of normalized-centered protein abundance of significantly differentially abundant proteins between strains. CtV+ samples are labeled in blue, and CtV- samples are labeled in orange. K: CtV+ venom samples; M: CtV- venom samples. Numbers correspond to the replicates. The five most relative abundant proteins are highlighted in red.

| Protein ID | Target | Species | e-value | Log10Fold | iBAQ | Excretion signal | QTL |
|---|---|---|---|---|---|---|---|
| COTY_00001603 | Similar to APN2: Aminopeptidase N (Manduca sexta) | *Cotesia congregata* | 0.00 | 1.504 | 1,346,288 | None | - |
| KAG8039124.1 | Serine protease inhibitor (SERPIN) | *Cotesia flavipes* | 1.00e-50 | 1.619 | 440,104 | Yes | 1-ON, 2-PS |
| KAG8039128.1 | Serpin B4 | *Cotesia chilonis* | 0.00 | 1.513 | 127,730 | Yes | 1-ON, 2-PS |
| COTY_00015768 | ATP-citrate synthase | *Cotesia chilonis* | 0.00 | 1.793 | 1,701 | None | - |
| COTY_00016350 | Protein disulfide-isomerase A6 homolog | *Cotesia glomerata* | 0.00 | -1.553 | 548,362 | Yes | - |
| KAG8035893.1 | Immunoevasive protein 2B | *Cotesia chilonis* | 0.00 | -2.242 | 80,193 | Yes | - |
| COTY_00008275 | Putative annexin B9-like isoform X1 | *Cotesia chilonis* | 0.00 | -2.222 | 62,203 | None | - |
| COTY_00012191 | Putative FK506-binding protein 2 | *Cotesia chilonis* | 8.00e-151 | -1.831 | 35,883 | None | - |
| KAG8041199.1 | Extracellular superoxide dismutase 3 | *Cotesia chilonis* | 2.00e-116 | -2.423 | 20,407 | Yes | - |
| COTY_00016584 | Similar to armt1: Damage-control phosphatase ARMT1 | *Cotesia congregata* | 0.00 | -1.960 | 1,391 | None | - |
| COTY_00006567 | Cysteine-rich with EGF-like domain protein 2 | *Cotesia glomerata* | 0.00 | -1.990 | 1,921 | Yes | 1-ON, 2-PS |
| COTY_00010445 | Uncharacterized protein LOC123263598 | *Cotesia glomerata* | 1.00e-143 | -1.503 | 1,896 | Yes | - |

Table 1: Blast identification of the 12 differentially abundant proteins between CtV+ and CtV-. Proteins were blasted against the nr NCBI database, and only the best hit (e-value and query cover) of each query was reported. Positive Log10Fold means over-abundance in CtV+. Hyphens mean no match in the QTLs.
Proteins of ID "COTY" were present in the new annotation, and proteins with ID "KAG" were found only in the former *C. typhae* proteome.

Eight proteins were over-abundant in the CtV- strain against four in the CtV+ strain (Table 1). Log10Fold changes were similar between proteins, ranging from 1.504 to 2.423, but proteins were more strongly over-abundant in CtV- than in CtV+. Most proteins were identified as putative venom by the iVenomDB and they were blasted against the nr NCBI database in order to confirm and precise their potential function. All 12 proteins were similar to known proteins from other species of the *Cotesia* genus. Eight proteins corresponded to newly annotated genes, and four to genes present only in the former annotation.

Proteins of ID COTY over-abundant in CtV+ were encoded by genes that were also over-expressed in CtV+ in the free stages of the wasp. No over-abundant protein in CtV- was encoded by a gene that was over-expressed in CtV- in the free stages. Two of CtV+ strain's over-abundant proteins were similar to serpins and belonged to the QTL 1-ON,2-PS, which is associated with a greater parasitism success of CtV+. Protein COTY_00001603, which was similar to an aminopeptidase, was far more abundant than the other 11 proteins, and thus represented a very interesting candidate protein. Moreover, it was the fourth most abundant protein in the venom.

### 3.2. RNA differential expression

Abdomen RNA samples were analyzed by differential expression analysis using the DESeq2 R package. Expression was analyzed with a model of type expression ~ strain*stage. Numbers of d.e. genes hereafter reported include globally d.e. genes and those specifically d.e. in one condition of the second factor.

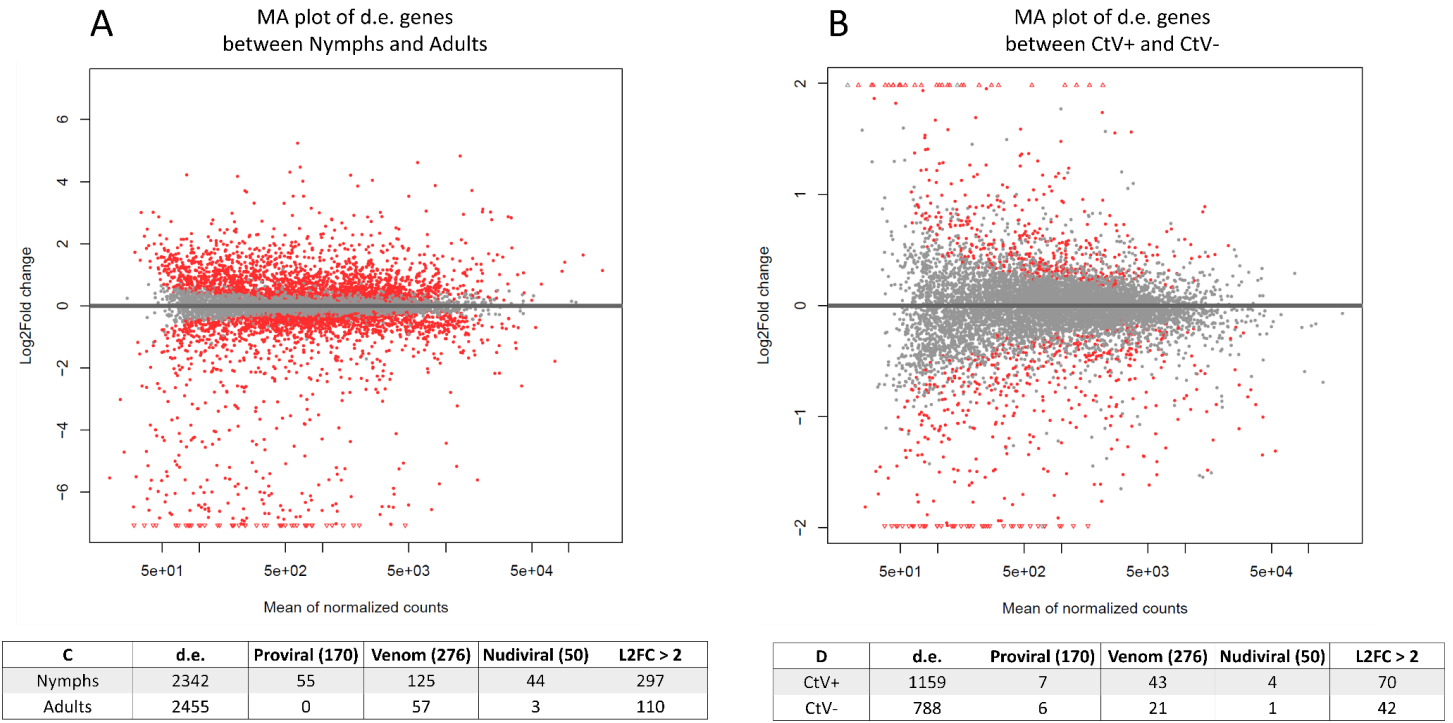

Figure 4: Differentially expressed genes in the FSE. (A) MA plot of over-expressed genes between life stages. Positive Log2Fold change means over-expression in adults. Significantly differentially expressed genes are labelled in red. (B) MA plot of over-expressed genes between strains. Positive Log2Fold change means over-expression in CtV-. Values are numbers of over-expressed genes (C) between life stages and (D) between *C. typhae* strain. First column reports the number of d.e. genes, second, third, fourth and fifth detail the number of d.e. genes that are proviral, venom, nudiviral and having a Log2Fold change > 2. Numbers in headers correspond to the number of genes in the genome. Venom genes correspond to putative venom genes (see § 3.1) that are present in the new annotation.

The repartition of over-expressed genes detected in the FSE showed that, consistent with the PCA analysis (Fig. 2, A), there were more d.e. genes between life stages (Fig. 4, A, C) than between strains (Fig. 4, B, D). The two strains still significantly differed, with CtV+ displaying more over-expressed genes than CtV- (Fig. 4, D), particularly more venom genes and more strongly over-expressed genes.

Even if there was no expression bias between life stages, there were more strongly over-expressed genes (with Log2Fold > 2) in nymphs than in adults (Fig. 4, A, C). Among the genes strongly over-expressed in nymphs were 53 proviral genes and 42 nudiviral genes. Venom genes were not strongly over-expressed in nymphs or adults.

### 3.2.1. Venom gene expression in the free stages of *C. typhae*

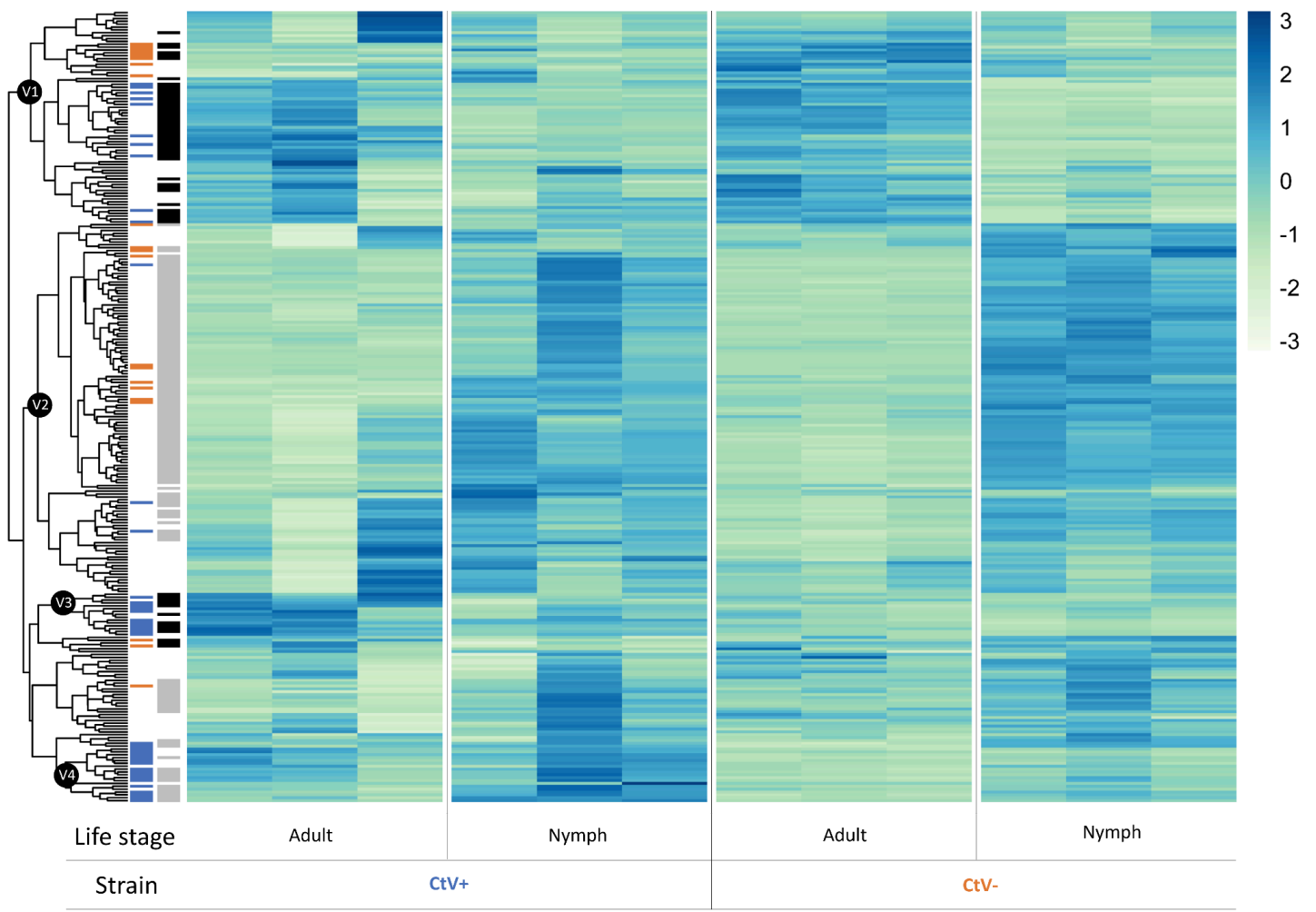

Figure 5: Heatmap of the normalized-centered expression in *C. typhae* free stages of the 276 genes identified by iVenomDB, SignalP and venom apparatus proteomic. Genes over-expressed in CtV+ strain are labeled in blue, and those in CtV- are labeled in orange. Genes over-expressed in nymphs are labeled in grey and those in adults are labeled in black. Four nodes of interest are selected (black numbers).

In order to identify candidate virulence genes, we focused on the 276 *C. typhae* newly annotated putative venom genes (see § 3.1).

The expression profile of those genes is shown in Fig. 5. Third replica of CtV+/Adult displayed a very different profile, which seems misleading but did not considerably affect the heatmap clustering.

Putative venom genes were mostly differentially expressed between life stage conditions, with a bias toward expression in nymphs (125 over-expressed genes, Fig. 4, C, and Fig. 5, grey, node V2) rather than in adults (57 over-expressed genes, Fig. 4, C and Fig. 5, green, node V1).

Regarding strains, more putative venom genes were over-expressed in CtV+ (43) than in CtV- (21), and over-expressed CtV+ genes clustered (Fig. 5, nodes V3 and V4) while those of CtV- did not, indicating a more consistent expression of CtV+ venom genes. Moreover, over-expressed CtV+ genes were segregated into two sub-clusters whether they were also over-expressed in adults (Fig. 5, node V3) or nymphs (Fig. 5, node V4).

### 3.2.2. Proviral and nudiviral expression in the free stages of *C. typhae*

In the free stages of *C. typhae*, we detected the expression of 156 out of 170 proviral genes, which was mostly limited to the nymph stage. No proviral genes were over-expressed in adults, while 55 were in nymphs, in both strains (Fig. 4, C). There was very little proviral differential expression between strains, with 7 genes over-expressed in CtV+ and 6 in CtV-. Interestingly, 39 proviral genes located on various segments were only expressed in the free stages (and not in the parasitized host), 6 of which were over-expressed in the nymphs. We could not detect any expression of 9 proviral genes, either in the free or the parasitic stage.

A bias of expression toward nympal stage was also observed for nudiviral genes, as 44 nudiviral genes were over-expressed in nymphs, against 3 in adults. Concerning the strains, there were very few differences, with 4 genes over-expressed in CtV+ and 1 in CtV-. Nudiviral genes over-expressed in CtV+ were similar to odve66_3, odve66_32, lef5 and HzNVorf64_a of *C. congregata*, and had Log2Fold change ranging from 0.52 to 4.16. The nudiviral gene over-expressed in CtV- was similar to odve66_3 of *C. congregata* and had a Log2Fold change of 0.99.

### 4. Parasitic stages expression

Hemolymph RNA samples were analyzed by differential expression analysis using the DESeq2 R package. First, we analyzed the expression data with a model of type expression ~ strain*host*time. Then, we split the dataset into two subsets, depending on time. Expression at 24 hours and 96 hours post-parasitism were then analyzed with models of type expression ~ strain*host. Numbers of differentially expressed (d.e.) genes hereafter reported include globally d.e. genes and those specifically d.e. in one condition of the second factor.

The whole parasitic-stage experiment was analyzed with DESeq2 in order to check for the differential expression between time conditions. While this analysis was clearly biased, given that only 54 genes are expressed at 24 hours, it could yield interesting insights into proviral gene expression patterns. We found 31 proviral genes over-expressed at 24 hours against 22 at 96 hours, showing that proviral expression was not constrained to the early stages of parasitism. Most of the genes over-expressed in one time condition belonged to the same circles (e.g. circles 1_Duplication, 2 and 7 at 24h, circles 12, 16 and 32 at 96h), possibly indicating that circle expression was constrained by time.
Out of the 170 proviral genes, 48 were not expressed at all in the two time conditions. They were equally distributed between the circles, but we reported that all genes of segment 5 and 6 (three and two genes, respectively), which do not integrate in host cells (Muller, Chebbi, et al., 2021), did not express in the host.

#### 4.1. Differential expression of the 54 *C. typhae* proviral genes expressed at both time points (24 and 96 hours)

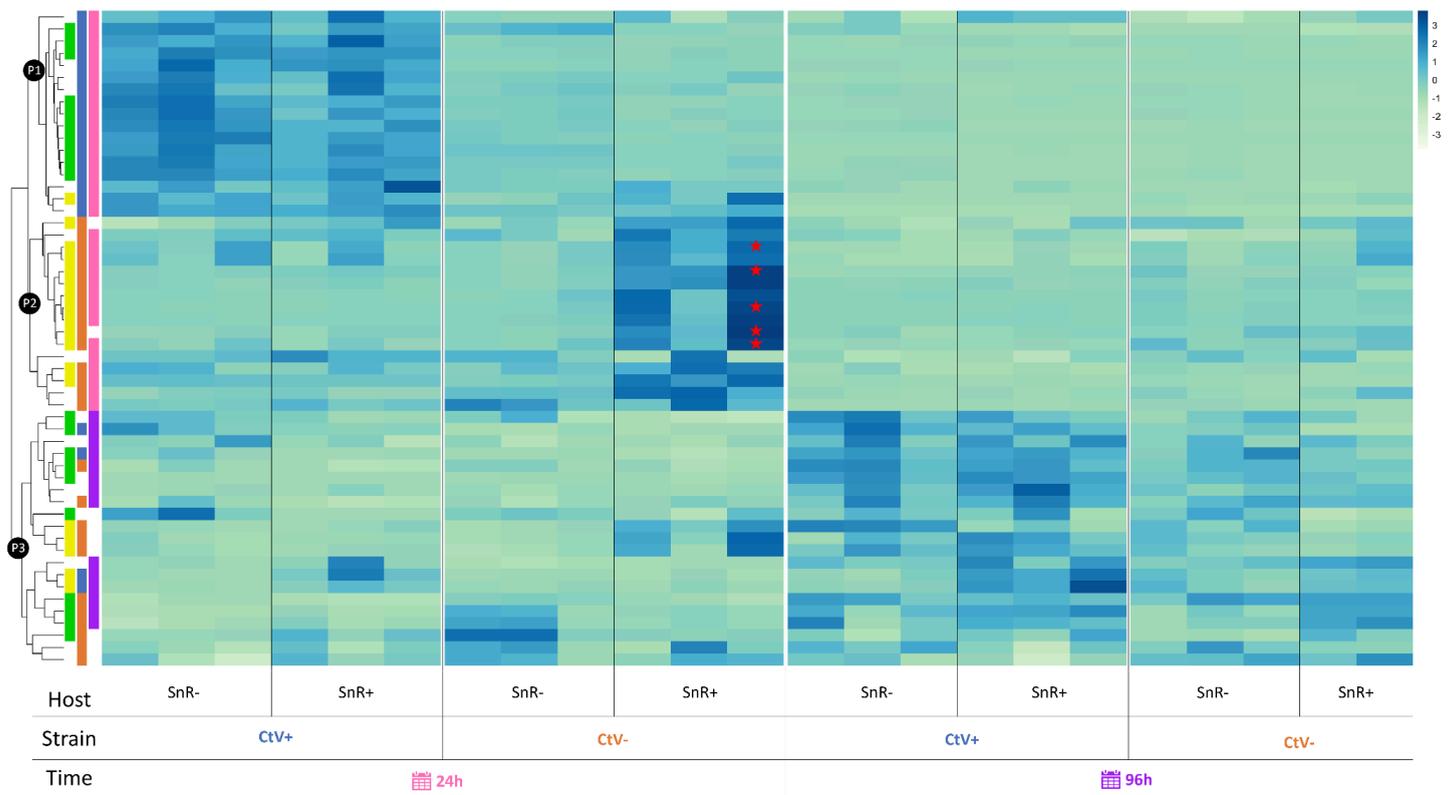

Figure 6: Heatmap of the normalized-centered expression of the 54 genes expressed at both 24 and 96 hours, depending on conditions. Genes over-expressed in specific conditions are labeled with colors: SnR- (green), SnR+ (yellow), CtV+ (blue), CtV- (orange), 24 hours (pink), and 96 hours (purple). Three nodes, P1, P2 and P3, are labelled. Genes from node P2 belonging to the 1-ON,2-PS QTL are labeled with red stars.

The 54 genes remaining after filtration for low counts in the 24-PSE were all proviral genes which were also expressed at 96h. Their expression pattern was analyzed through a heatmap (Fig. 6). Genes mostly clustered according to differential expression between time conditions (Fig. 6, pink and purple).

Regarding differential gene expression 24 hours post-parasitism, and consistent with the PCA analysis (Fig. 2, B), the greatest number of differentially expressed (d.e.) genes was found between the parasitoid strains (21 over-expressed in CtV+, and 27 in CtV-) (Fig. 6, blue and orange). Moreover, differential expression between strains separated the genes over-expressed at 24h into two distinct nodes, P1 (genes over-expressed in CtV+) and P2 (genes over-expressed in CtV-). Gene expression also varied between host populations, with 20 genes over-expressed in SnR- and 18 in SnR+ (Fig. 6, green and yellow).

Among the genes over-expressed at 24h, all genes over-expressed in SnR- were also over-expressed in CtV+ (Fig. 6, node P1) and almost all genes over-expressed in SnR+ were also over-expressed in CtV- (Fig. 6, node P2). While expression of CtV+ genes was very constant between host populations at 24 hours, that of CtV- varied considerably more, indicating a strong effect of the host population on CtV- early proviral expression. Thus, expression differences between host populations were exacerbated in the CtV- samples (33 d.e. genes) compared to the CtV+ samples (4 d.e. genes). In CtV- samples, genes were more strongly over-expressed in SnR+ (12 genes with Log2Fold > 1.5) than in SnR- (0 genes with Log2Fold > 1.5). Genes of node P2 were mostly over-expressed in SnR+ hosts parasitized by CtV-. Five of those genes (Fig. 6, red stars) belong to the QTL 1-ON,2-PS (Benoist, Capdevielle-Dulac, et al., 2020), which is associated with a weaker parasitism success for the CtV- strain. Four of them were identified as protein tyrosine phosphatase, and one as EP1-like protein.

Consistent with what Benoist and collaborators (2017) found, both CrV1 and Cystatin proviral genes were over-expressed in CtV+ at 24 hours post-oviposition. However, differential expression of CrV1 was not detected when restraining analysis to SnR-, potentially indicating that host population has an impact on CtV- CrV1 expression.

### 4.2. Differential expression of the 5,245 *C. typhae* genes expressed at 96 hours

| A | d.e. | Proviral (170) | Venom (276) | L2FC > 2 | B | d.e. | Proviral (170) | Venom (276) | L2FC > 2 |
|---|---|---|---|---|---|---|---|---|---|
| CtV+ | 524 | 3 | 18 | 43 | SnR- | 17 | 6 | 2 | 6 |
| CtV- | 537 | 13 | 20 | 50 | SnR+ | 15 | 0 | 1 | 5 |

Table 2: Differential expression of all *C. typhae* genes expressed in the 96-PSE. Values are the number of over-expressed genes (A) between *C. typhae* strains and (B) between host populations. First column reports the number of d.e. genes, second, third and fourth detail the number of d.e. genes that are proviral, venom, and having a Log2Fold change > 2. Numbers in headers correspond to the number of genes in the genome. Venom genes correspond to putative venom genes (see § 3.1) that are present in the new annotation.

Consistent with the PCA analysis (Fig. 2), there were more d.e. genes between parasitoid strains (Table 2, A) than between host populations (Table 2, B), which tended toward more genes over-expressed in CtV-, especially for proviral genes.

Unlike expression at 24h, there were only 3 d.e. genes between host populations in the CtV- strain while there were 14 in the CtV+ strain, indicating that the host population effect on CtV- gene expression disappeared with time.

Out of the 537 genes over-expressed in the CtV- strain, 23 belonged to the QTL 1-ON,2-PS, of which 7 were proviral genes (located on segment 1) and 5 were those identified in the 24-PSE (Fig. 6). For comparison, only 4 genes from this QTL were over-expressed in CtV+. Thus, over-expression of genes from this QTL in the CtV- strain lasted over time. However, the pattern observed in the 24-PSE was no longer visible at 96 hours, as no genes from this QTL were differentially expressed between host populations when focusing on the CtV- strain.

### 5. Candidate genes identified through combined approaches

Combining transcriptomic and proteomic approaches crossed with QTL data, lead to selection of few candidate genes (Table 3) that could play a key role in the virulence difference between *C. typhae* strains. Genes considered as candidates were either detected by combined approaches or displaying outlier features (very high protein abundance or Log2Fold change in differential expression).

| Gene | Feature(s) | Over-expression | | | Over-abundance | Cluster | Annotation |
|---|---|---|---|---|---|---|---|
| | | FSE | 24-PSE | 96-PSE | | | |
| COTY_00006518 | Proviral QTL 1-ON,2-PS | CtV-, at nymph stage (L2F = 1.400) | CtV- (L2F = 1.426) | CtV- (L2F = 1.652) | - | 5 | Protein tyrosine phosphatase (*C. vestalis bracovirus*) e-value = $10^{-66}$ |
| COTY_00003145 | iVenomDB Most over-epxressed | CtV+ (L2F = 6.425) Nymphs (L2F = 0.735) | - | - | - | 21 | Putative uncharacterized protein (*C. vestalis*) e-value = $10^{-71}$ |
| COTY_00009342 | Putative venom QTL 2-ON | CtV+ (L2F = 0.995) Nymphs (L2F = 0.952) | - | - | - | 21 | Venom metalloprotease (*M. demolitor*) e-value = $10^{-8}$ |
| COTY_00016350 | Putative venom Abundant in venom | Nymphs (L2F = 1,390) | - | CtV+ (L2F = 0,693) | CtV- (L10F = 1.553) | 9 | Protein disulfide-isomerase A6 homolog (*C. glomerata*) e-value = 0.00 |
| KAG8035893.1 | Putative venom Abundant in venom | - | - | - | CtV- (L10F = 2,242) | - | Immunoevasive protein 2B (*C. chilonis*) e-value = 0.00 |
| COTY_00001603 | Putative venom Abundant in venom | CtV+ (L2F = 1,211) Nymphs (L2F = 0,741) | - | - | CtV+ (L10F = 1.504) | 3 | APN2 : aminopeptidase N (*M. sexta, C. congregata*) e-value = 0.00 |
| KAG8039124.1 | Putative venom Abundant in venom QTL 1-ON,2-PS | - | - | - | CtV+ (L10F = 1.619) | - | Serpin (*C. flavipes*) e-value = 0.00 |
| KAG8039128.1 | Putative venom Abundant in venom QTL 1-ON,2-PS | - | - | - | CtV+ (L10F = 1.513) | - | Serpin B4 (*C. chilonis*) e-value = $10^{-50}$ |

Table 3: Candidate genes selected through combination of all approaches used in the experiment. Features indicate genomic and proteomic data, such as presence in a QTL, high protein abundance, or identification as venom. L2F reports the RNA Log2Fold change when significant, and in the condition of over-expression. L10F reports the proteic Log10Fold change, in the condition of over-abundance. Cluster belonging (see § 6) is reported. Hyphens mean no results or no data.

Gene COTY_00006518 was one of the three proviral genes over-expressed in CtV- specifically in nymphs, which belonged to the node P2 containing genes over-expressed at 24 hours in CtV- in the SnR+ host (Fig. 6). When restraining the differential expression analysis to SnR-, this gene was not detected as d.e. (Log2Fold change = 0.508, p-adj = 0.327), which tended to show that over-expression of this gene was actually constrained to the SnR+ host population, consistent with the other genes of its cluster (Fig. 6). At 96h, this gene is still globally over-expressed in CtV-, but the fold change is higher when focusing on SnR+ samples (Log2Fold change = 2.162, p-adj < 0.001), indicating that it is still more over-expressed in CtV- parasitizing SnR+ than SnR-. Thus, it represented a potential candidate gene for explaining virulence differences between strains, as its expression patterns seemed to be specific to the forbidding interaction between CtV- (less virulent) and SnR+ (more resistant).

Gene COTY_00003145 was the most over-expressed CtV+ gene that could be identified in the FSE. It was identified by the iVenomDB as similar to a venom protein of *C. vestalis* which function is not identified yet. However, the proteic sequence matches several chemosensory proteins through blastp against all NCBI databases, in several Lepidopteran species of the taxa Ditrysia, some Hemipteran (such as *Matsumurasca*) and some Pscocodea (such as *Liposcelis*).

Among the putative venom genes (see § 3.1) that were d.e. between CtV+ and CtV- in the FSE, only COTY_00009342 belonged to a QTL.

Five other genes were selected as candidates based on the differential abundance of their protein between strains and their high relative abundance.

### 6. Clustering

Genes were clustered with HTSCluster according to their expression profiles in the FSE and in the 96-PSE. Due to the bias introduced by the very low number of expressed genes in the 24-PSE (54 genes expressed), those conditions were removed from this analysis. Genes were beforehand filtered by HTSFilter, which conserved 8,751 genes, including 68 proviral, 47 nudiviral and 270 putative venom genes. After clustering, we chose the structure made by DDSE algorithm, as BIC and ICL converged towards a too high number of clusters to allow for subsequent analysis. The analysis yielded 45 clusters that contained between 86.67 and 100.00% of genes with probability of belonging > 0.95, representing between 32 and 644 genes per cluster.

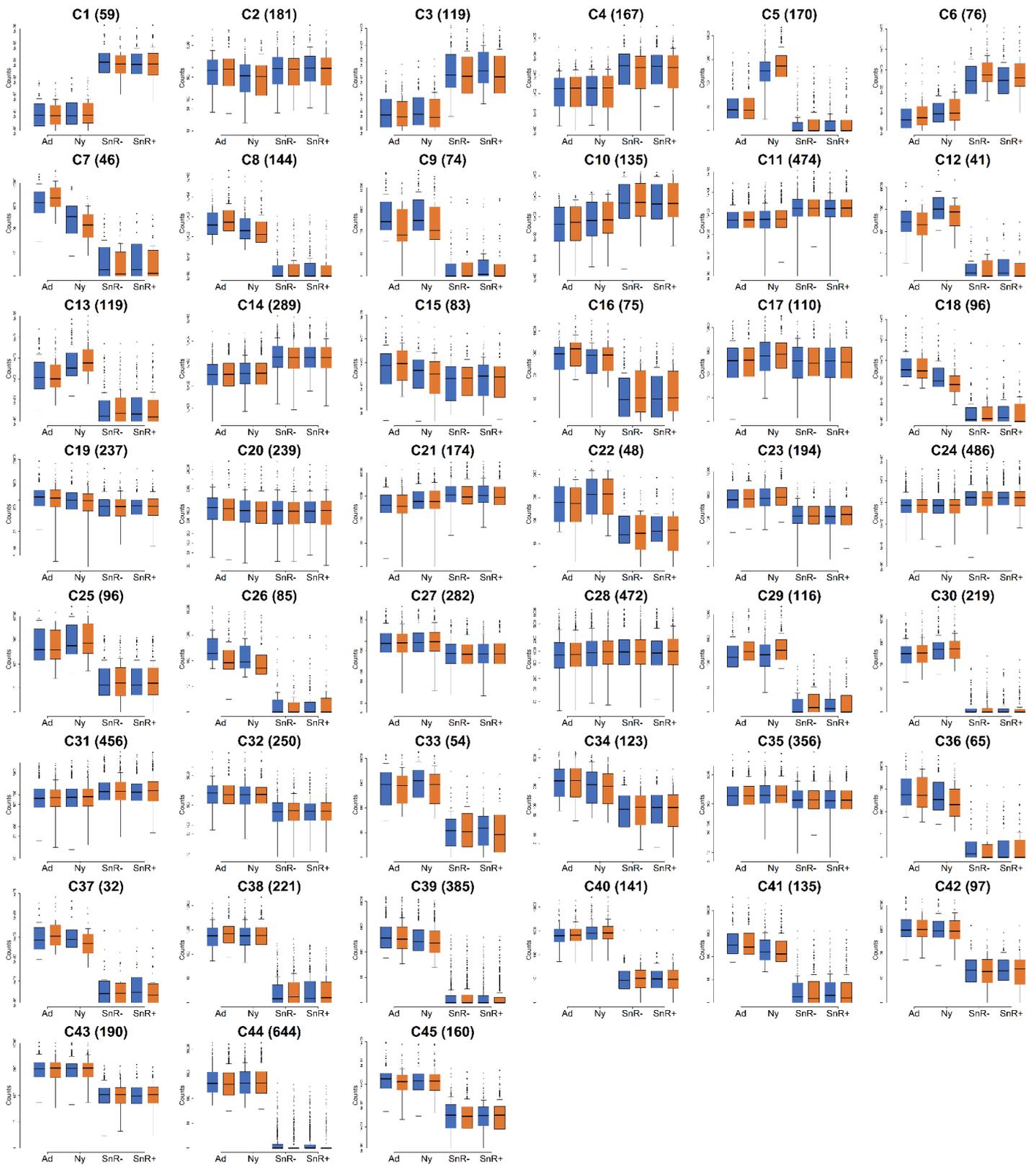

Figure 7: Distribution of gene expression across RNA-seq conditions, for the 45 clusters detected by DDSE algorithm of HTSCLuster. Counts were log-transformed after normalization by DESeq2. Four first boxes are expression in the free stages, four last are expression in the host, 96 hours post-parasitism. Box colors: blue: CtV+ strain; orange: CtV- strain. Labels: Ad: adult life stage; Ny: nymph life stage; SnR-: permissive host; SnR+: resistant host.

Clustering was mostly driven by expression profiles between the FSE and the 96-PSE (Fig. 7). Many clusters showed a clear differential expression between adults and nymphs (clusters 2, 5, 7, 8, 12, 13, 15, 18, 34, 36, and 41). Few were driven by differential expression between CtV+ and CtV- in the free stages. For example, clusters 9, 18 and 26 contained mostly genes over-expressed in CtV+, and cluster 16 and 29 genes over-expressed in CtV-. Similarly, 54 of the 119 genes of cluster 3 were over-expressed in CtV+ in the 96-PSE, and clusters 6 and 34 contained a majority of genes over-expressed in CtV- in this experiment.

The 68 proviral genes were present in 14 clusters, but were more concentrated in clusters 5 and 6, which contained respectively 10.0% (17 genes) and 19.4% (15 genes) of proviral genes. Cluster 6 also contained 8 out of 12 genes of node P2 identified in § 2.2.1. Globally, clusters 5, 6 and 10 contained 58.8% of the 68 proviral genes.

The 47 nudiviral genes were almost all located in cluster 5 (41 genes, i.e. 24.1% of the cluster). The 6 nudiviral genes left were dispersed in 6 different clusters, and comprised the 3 genes over-expressed in adults. The 270 putative venom genes were also evenly dispersed in 42 clusters, with clusters 4, 9 and 10 containing respectively 11.3% (19), 13.0% (10 genes) and 14.7% (20 genes) of venom genes.

Among selected candidate genes, COTY_00003145 and COTY_00009342 were located in cluster 21, which was composed of 86.8% of genes over-expressed in nymphs. COTY_00006518 was located in cluster 5, which was also composed of genes over-expressed in nymphs, but also contained 12.9% of genes that were over-expressed in CtV- at the nymph stage.

The candidate protein over-abundant in CtV+, COTY_00001603, belonged to cluster 9 which was composed by 93.2% of genes over-expressed in CtV+ in the FSE. The candidate protein over-abundant in CtV-, COTY_00016350, belonged to cluster 3 which was characterized by over-expression of CtV+ in the 96-PSE (45.4% of genes).

7. Enrichment analysis

7.1. Cluster enrichment analysis

When performing GO enrichment analysis on the 45 clusters, TopGO was the least stringent method to find enriched GO terms, yielding 3 220 enrichments over all clusters. The find_enrichment function from GOATools yielded 1974 enrichments, and the ClusterProfiler package remained the most stringent, yielding 506 enrichments. Overall, 96 GO terms were found to be enriched by all three methods, in 14 different clusters (Table 4).

| Cluster | Category | GO ID(s) | Function(s) | Cluster content |
|---|---|---|---|---|
| 4 | BP | GO:0018279 ; GO:0034975 ; GO:0006890 ; GO:0007030 ; GO:0048194 ; GO:0030968 ; GO:0006888 | Protein maturation, Golgi activity, endoplasmic reticulum activity | 11.3% putative venom genes<br>31.7% o.-e. in nymphs<br>26.9% o.-e. in CtV+ parasitizing SnR-, 96-PSE |
| | CC | GO:0008250 ; GO:0005789 ; GO:0033116 ; GO:0000139 ; GO:0098791 ; GO:0005790 ; GO:0005788 ; GO:0030133 ; GO:0005793 ; GO:0030660 ; GO:0030658 | Protein transfert, Golgi and endoplasmic reticulum membrane and lumen | |
| | MF | GO:0030246 | Carbohydrate binding | |
| 5 | MF | GO:0008010 | Structural constituent of chitin-based larval cuticle | 10.0% proviral genes<br>24.1% nudiviral genes<br>99.4% o.-e. in nymphs<br>12.9% o.-e. in CtV-, nymph stage |
| 6 | BP | GO:0021898 | Commitment of multipotent stem cells to neuronal lineage in forebrain | 19.4% proviral genes<br>38.2% o.-e. in CtV- parasitizing SnR-, 96-PSE |
| 10 | BP | GO:0002181 ; GO:0006614 ; GO:0006413 ; GO:0000184 ; GO:0097421 ; GO:1904587 ; GO:0000027 ; GO:0042694 ; GO:0002726 ; GO:0042273 ; GO:0006446 ; GO:0006364 | Translation, RNA processing, liver regeneration, response to glycoprotein, muscle cell fate specification, positive regulation of T cell cytokine production | 14.7% putative venom genes<br>17.0% o.-e. in CtV- parasitizing SnR-, 96-PSE<br>60.0% o.-e. in nymphs |
| | CC | GO:0022625 ; GO:0022627 ; GO:0042788 ; GO:0098556 ; GO:0005844 | Ribosome, polysome | |
| | MF | GO:0003735 ; GO:0070180 ; GO:0043024 ; GO:0031369 ; GO:0019843 ; GO:0003729 | Ribosome binding, translation | |
| 11 | BP | GO:0106074 ; GO:0048194 ; GO:0006418 | Translation, Golgi activity | 30.0% o.-e. in nymphs<br>11.2% o.-e. in adults<br>11.8% o.-e. in CtV- parasitizing SnR-, 96-PSE |
| | CC | GO:0005730 ; GO:0005759 | Nucleolus, mitochondrial matrix | |
| | MF | GO:0004812 ; GO:0000049 ; GO:0002161 | tRNA activity | |
| 13 | BP | GO:0006030 | Chitin metabolic process | 21.5% o.-e. in CtV-, FSE<br>16.0% o.-e. in CtV-, nymph stage<br>98.3% o.-e. in nymphs |
| | MF | GO:0008061 | Chitin binding | |
| 14 | BP | GO:0002181 ; GO:0000184 ; GO:0006413 ; GO:0006614 | Translation | 35.3% o.-e. in nymphs<br>12.1% o.-e. in CtV+ parasitizing SnR-, 96-PSE |
| | CC | GO:0022625 ; GO:0005789 | Ribosome, endoplasmic reticulum | |
| | MF | GO:0003735 | Structural constituent of ribosome | |
| 17 | BP | GO:0009167 ; GO:0009205 ; GO:0009150 ; GO:0046496 ; GO:0006734 | Nucleotide, NADH metabolic process | 13.6% o.-e. in CtV-, FSE<br>98.2% o.-e. in nymphs |
| | CC | GO:0098803 ; GO:1990204 ; GO:0005746 ; GO:0070469 ; GO:0098798 | Mitochondrial respiration | |
| | MF | GO:0009055 | Electron transfer activity | |
| 19 | BP | GO:0006338 ; GO:0008361 | Chromatin remodeling, regulation of cell size | 71.7% o.-e. in adults |
| | MF | GO:0005085 ; GO:0051020 | Nucleotide | |
| 20 | BP | GO:0045814 | Negative regulation of gene expression, epigenetic | 14.2% o.-e. in CtV+, FSE<br>55.6% o.-e. in adults<br>10.8% o.-e. in adults, CtV+ strain |
| 21 | BP | GO:0009205 ; GO:0006120 ; GO:0009167 ; GO:0032981 ; GO:0009150 | Nucleotide, mitochondrial respiration | 13.2% o.-e. in CtV+<br>86.8% o.-e. in nymphs<br>15.5% o.-e. in CtV+ parasitizing SnR-, 96-PSE |
| | CC | GO:0005747 ; GO: 0043209 | Mitochondrial respiration, myelin sheath | |
| | MF | GO:0009055 | Electron transfer activity | |
| 24 | BP | GO:0016579 ; GO:0043161 ; GO:0006886 | Protein transport, ubiquitination | 21.3% o.-e. in adults |
| | CC | GO:0005770 ; GO:0045334 ; GO:0005802 | Endosome, endocytosis, Golgi | |
| | MF | GO:0043021 | Ribonucleoprotein complex binding | |
| 27 | BP | GO:0051053 ; GO:0006260 | Replication | 20.7% o.-e. in nymphs |
| | CC | GO:0030894 | Replisome | |
| 45 | CC | GO:0051233 | Spindle midzone | 11.9% o.-e. in CtV+, FSE<br>20.0% o.-e. in Kobobo, adult stage<br>13.8% o.-e. in adults, CtV+ strain |

Table 4: Enriched GO terms found in the FSE and 96-PSE gene expression clusters. GOs with similar functions were regrouped under broader terms. BP: Biological process, CC: Cellular components, MF: Molecular function. Cluster content represents percentages of genes of each cluster belonging either to a gene category or being over-expressed (o.-e.) in a certain condition.

Among clusters with enriched GO terms, clusters 5, 13, 17 and 21 contained mostly genes over-expressed in nymphs. They were enriched with terms related to chitin metabolism and mitochondrial respiration. Clusters 19 and 20 mostly contained genes over-expressed in adults and terms related to gene expression regulation and cell cycle. Cluster 45 had 18.13% of its genes over-expressed in CtV+ when focusing on adult condition, and only had one term enriched, related to cell mitosis. The other clusters did not display a clear link between GO enrichment and transcriptional profile.

### 7.2. Gene subset enrichment analysis

Genes that were over-expressed in CtV+ in the FSE were determined by the three methods to be enriched with one term, GO:0022617, related to extracellular matrix disassembly. Genes over-expressed in CtV- in the FSE were determined by the three methods to be enriched with two terms, GO:0006030 and GO:0008061, related to chitin metabolic process and chitin binding. No enriched terms were found in genes over-expressed in CtV+ or CtV-, 96 hours post-parasitism.

Putative venom genes were enriched with 48 GO terms found with the three methods, mostly involved in molecular binding (11 terms), cell shape (11 terms), protein folding and cell integrity (7 terms), metabolic activity (6 terms) and enzyme activity (6 terms).

Proviral genes were enriched with only two GO terms in the three methods, GO:0035335 and GO:0004725, related to peptidyl-tyrosine dephosphorylation and protein tyrosine phosphatase activity. Both terms are involved in cleaving phosphate groups from tyrosine amino acids of target proteins.

# Discussion

This study aimed at characterizing, through combined approaches of transcriptomic and proteomic, the molecular basis of *Cotesia typhae* virulence and its differences between two strains, one being highly virulent (CtV+) and the other being weakly virulent (CtV-). Those strains display a parasitism success that is similar on the permissive host population (SnR-) and very different on the resistant host population (SnR+). Overall, we found that virulence factors were preferentially expressed at the nymph stage, with some differences between CtV+ and CtV-. Venom composition appeared to be slightly different, with few key proteins varying in abundance between strains. Expression in the host revealed a clear differentiation in proviral gene expression between strains, modulated by the host population in the case of CtV- strain.

1. Virulence factor expression dynamic

Clustering on genes expressed in the free stages of *C. typhae* and 96 hours post-parasitism allowed to globally analyze virulence factor expression dynamics at those stages. First, clustering showed that differential expression between adult and nymph stages was the second main factor grouping genes together, after the difference between free and parasitic stages. This is consistent with the very high number of genes that were d.e. between adults and nymphs. Division into several clusters was likely the result of different expression patterns in the remaining conditions. Then, by analyzing differential gene expression at those stages, we aimed at identifying temporal structure in proviral, nudiviral or venom genes expression. Finally, most enriched GO terms identified in clusters were linked to insect metamorphosis, mitochondrial activity, protein translation, and post-traductionnal modification, which could be linked to the high number of genes over-expressed in nymphs found in most of the clusters.

1.1. Proviral gene expression

Surprisingly, we found strong expression of proviral genes in the free stages, which were mostly over-expressed in nymphs. Very little is documented about proviral expression in nymphs and adults, although it has been shown for *ptp* and *ank* genes of *Microplitis demolitor* (Bitra et al., 2011). Virulence genes such as *ptp*, *serrich* and *Egf* were also expressed in free stages of *Cotesia typhae*, therefore this expression is not only due to viral segments amplification and replication. However, as pointed out by Bitra and collaborators (2011), we cannot tell if those transcripts are actively translated or if this expression is constrained to female calyx. We can hypothesize that if the virulence proteins are produced, they can be either packaged in the virions or participate in the ovarian fluid that is injected along the eggs, and therefore act as immunosuppressors in the very early stages of parasitism, as venom can (Gauthier, Drezen, et al., 2018). Some of those genes appear to be only expressed in the free stages, which would therefore indicate that they are translated into virulence factors outside of the host. They could also have a very short expression kinetic not encompassed by the two time points used in this experiment. Moreover, we found a small number of proviral genes that were not expressed at all in any conditions. Either their expression pattern is very short, as previously hypothesized, or they constitute pseudogenes, as several exist within the *C. typhae* genome, probably as the consequence of host specialization (Gauthier, Gayral, et al., 2018). If so, they should be at the very first stages of pseudogenisation, as they were still annotated and identified as similar to bracoviral genes, thus not displaying important deletion or nonsense mutation.

In the host, only 54 genes, all proviral, were expressed 24 hours post-parasitism, against 5,245 at 96 hours. This increase in the number genes expressed and their diversity is likely due to expression in the teratocytes that have been liberated at this time point (Gornard et al., 2024), but also to late proviral expression. Delayed expression of some proviral genes has also been described in other parasitoid species as for example, protease inhibitor Egf0.4a that targets the phenoloxidase cascade to inhibit melanization of *Microplitis demolitor* (Lu et al., 2010; Strand, 2012b). Venom proteins can be quickly degraded inside the host, and immunosuppression can thus resort to proviral expression in the late stages of parasitism (Asgari, 2012).

Proviral genes were scattered among several clusters, consistent with their observed expression variability. However, more than 50% of those genes were concentrated in three clusters, mostly driven by nymph expression. Interestingly, the structured temporal expression of proviral genes, which could be expected (Strand, 2012a), seemed to be associated with the segment they belong to, with some being early expressed and others lately. Evolution could have favored synchronous expression of genes located on the same segment as a result of integration in host cells, but this association concerns only 53 of the 131 proviral genes expressed at those two time points, and therefore needs to be further investigated. Interestingly, differentially expressed proviral genes located on segments that do not integrate into the host genome were all over-expressed at 24h (Muller, Chebbi, et al., 2021). This possibly indicates that non-integrated segments need to be expressed early, even if they can maintain for up to 7 days in host tissues.

### 1.2. Nudiviral genes expression

Strong nudiviral gene expression was reported in the nymphs, as they are expressed toward the end of this life stage to induce proviral replication and virion production (Gauthier et al., 2021; Strand, 2012a). Our results confirm that this step is mostly constrained to the nymphal stage, as only three nudiviral genes are overexpressed in adults, involved in DNA replication and transcription. Nudiviral genes were almost all located in cluster 6, with the exception of 6 genes, among which those that were over-expressed in adults. Moreover, this cluster shows very little expression in the host, consistent with the absence of strong nudiviral expression in those conditions. Those results also indicate that DDSE clustering was relevant and reflected broad gene expression patterns.

### 1.3. Venom genes expression

Similar to nudiviral and proviral genes, putative venom genes expression seemed to occur preferentially in nymphs. *C. typhae* has a short adult lifespan (about 7 days when supplied with water and honey) and females mate when emerging from their cocoons, after which they start ovipositing (Kaiser et al., 2017). Our results are then consistent with the necessity for a functional virulence arsenal upon emergence, explaining the over-expression of virulence genes in nymphs along with highest Log2Fold changes. Putative venom genes were dispersed in several clusters, likely due to their high number and the high variability of their expression and role between conditions. Clusters 4 and 10 presented a high protein synthesis and maturation profile, which could be linked to the presence of several putative venom genes inside. Thus, genes with similar expression patterns, over-expressed in nymphs, could be involved in venom synthesis and post-traductional modification.

## 2. Virulence differences between strains

Gene clustering showed that differential expression between CtV+ and CtV- in the free stages or in the host drove the gene content of less clusters than the other conditions, according to the lower number of genes d.e. between the strains. By analyzing differential expression between strains and comparing their venom composition, we aimed at identifying molecular candidates involved in the virulence difference, whether in proviral, nudiviral or venom genes.

Overall in the free stages, CtV+ strain exhibited more over-expressed genes than CtV- with higher Log2Fold change, indicating a globally more active expression activity. However, almost all of the genes that were highly over-expressed in CtV+ (Log2Fold change > 2) corresponded to uncharacterized proteins, which were found neither in the QTL nor in the proviral segments. They were also not retrieved in the venom apparatus by proteomic approach, and do not correspond to potential venom proteins. This lack of overlap with known virulence factors is notably the case for genes included in cluster 45 that contain genes over-expressed in CtV+ strain in the adult stage. Interestingly, this cluster was associated with a term linked to cytoskeleton activity. As immunosuppression can be mediated by inhibition of hemocyte spreading, which resorts on cytoskeleton (Amaya et al., 2005; Richards & Edwards, 2002), cluster 45 could reveal production of hemocyte-disrupting virulence factors by CtV+ strain only. Genes over-expressed in CtV+ and for which no function was identified could correspond to novel virulence genes, such as ovarian protein genes. This suggests the necessity of a proteomic sequencing of *C. typhae* ovaries in order to detect novel virulence genes. This dataset could also be compared to the proteomic of the venom apparatus to detect functional overlap or tissue-specificity of putative venom genes expression.

Cluster 6 comprised one fifth of the proviral genes, mostly expressed in the host, and was enriched with a term associated with brain development. As *Cotesia* species are able to manipulate the nervous system of their host (Kelly et al., 1998; Shi et al., 2015), such a term could be associated with virulence factors encoded by proviral genes that are broadly present in this cluster. Interestingly, this cluster also contained a great amount of genes over-expressed in CtV-parasitizing SnR- host, perhaps indicating two different proviral virulence mechanisms used by this strain between the two hosts.

Finally, no pathway changes were revealed by enrichment analysis on d.e. genes between strains in the FSE and the 96-PSE. This could indicate that virulence involves several different functions that therefore cannot be detected with such analysis, consistent with the great number of d.e. genes between strains. Alternatively, this could also suggest that virulence relies on a few key genes rather than on a global virulence deficiency of the CtV- strain, which would be consistent with the similar parasitism success of both strains on the SnR- hosts, and would hint towards a gene for gene hypothesis, that we will further develop.

### 2.1. Nudiviral genes expression

In the free stages, the few nudiviral genes differentially expressed between strains were more numerous and more strongly over-expressed in CtV+ than in CtV- strain. As nudiviral genes are responsible for virion production and proviral packaging (Bézier et al., 2009; Gauthier, Drezen, et al., 2018), they are key factors for successful parasitism. CtV+ over-expressed nudiviral genes are involved in transcription (lef5, Strand & Burke, 2015) and capside construction (HzNVorf64 and odve_66, Arvin et al., 2021; Drezen et al., 2012). Odve_66 has been hypothesized to be involved in infectivity (Gauthier, Gayral, et al., 2018), allowing capsides to enter host cells, leading to proviral virulence gene expression. If CtV- virions lack essential surface factors to enter SnR+ cells, this could explain its weaker virulence on this host population.

### 2.2. Proviral genes expression

Similar to nudiviral genes, proviral ones showed very little expression differences between strains in the free stages. However, differential expression between strains in the early stages of parasitism was expected as it had previously been shown (Benoist et al., 2017). The approach used here allowed for a greater picture of early proviral expression and showed that the host population had more impact on CtV- than on CtV+. We could detect a group of genes (Fig. 6, P2) that were specifically over-expressed in CtV- parasitizing SnR+ and that contained five genes belonging to the QTL 1-ON, 2-PS, associated with a weaker parasitism success of CtV-. Four of those genes, including the candidate COTY_0006518, were similar to protein tyrosine phosphatase (*ptp*), a conserved gene family in bracoviruses. This family is found in several Microgastrinae (Provost et al., 2004), where they are thought to play a role in signal transduction and therefore in cellular immunity, which ensures encapsulation. The expression of *ptp* genes can arise early in parasitism (Gundersen-Rindal & Pedroni, 2006), and it can prevent metamorphosis or disrupt cellular response in *Spodoptera exigua* parasitized by *Cotesia plutellae* (Ibrahim & Kim, 2006; Kim et al., 2013). The last QTL gene of node P2 was similar to an EP1-like protein. EP genes are Early Expressed in parasitism and could represent an important source of host adaptation in *Cotesia* genus (Gauthier, Gayral, et al., 2018; Jancek et al., 2013). Expression level of *C. congregata* EP1 could be associated with higher *Manduca sexta* host susceptibility (Harwood et al., 1998). In hosts parasitized by *C. plutellae* and *C. congregata*, EP1 starts to be expressed 24 hours after oviposition (Chevignon et al., 2014; Lee et al., 2005). Furthermore, EP1-like proteins could have an immunosuppressive effect by diminishing hemocyte population (Kwon & Kim, 2008). CtV- strain does not have a hemolytic effect on SnR+, neither at 24 nor at 96 hours post-parasitism (Gornard et al., 2024, submitted). Therefore, we hypothesize that over-expression of those genes could be poorly efficient, allowing partial resistance of the host. Under-expression of proviral genes can explain parasitism failure on unsuitable hosts, as it is the case with *Microplitis demolitor* parasitizing *Trichoplusia ni* (Bitra et al., 2016). Here, we report over-expression in the case of parasitism failure, showing that gene regulation is more complex than expected. As the host population seems to have a great impact on CtV- proviral gene expression, it is possible that the resistant strain modulates this expression via retroaction loops.

At 96 hours post parasitism, host population seemed to no longer have an impact on strain expression, but the QTL genes of node P1 were still over-expressed in CtV-, even if this was not constrained to SnR+ population, showing that the virulence difference is mostly at play in the early stages of parasitism. Similar to proviral expression in the free stages, there were few d.e. genes between strains at 96h, indicating that differentiation between strains is mostly constrained to the early stages of parasitism.

### 2.3. Venom expression and abundance

Putative venom genes seemed to be globally more over-expressed in CtV+ than in CtV- strain and displayed a more consistent expression profile in CtV+ (Fig. 5), possibly causing a globally more concentrated venom in the CtV+ strain. Thus, if the SnR- population is more susceptible than SnR+ one, CtV- venom could be efficient enough to overcome its immune system, but not that of SnR+, while CtV+ venom would be efficient on both host populations. Concerning protein content, even if there were more over-abundant proteins in CtV- than in CtV+ strain, those of CtV+ were far more globally abundant than those of CtV-.

Significant differences in venom composition have already been reported between close strains of *T. brontispae* (Tang et al., 2019) or spatially scattered populations of *Leptopilina boulardi* (Mathé-Hubert et al., 2019). In general, venom composition can rapidly evolve through host adaptation (Poirié et al., 2014), independently of phylogeny. *C. typhae* strains score good parasitism success on SnR- population (CtV- natural host), but CtV- strain fails to successfully parasitize the natural host population of CtV+ (Benoist, Paquet, et al., 2020). As host-parasitoid interactions seems to be the main factor driving population divergence of sister species *C. sesamiae* (Branca et al., 2017), venom dissimilarity could be expected for *C. typhae* strains. Therefore, protein content of the whole venom apparatus of *C. typhae* was sequenced in order to identify putative venom genes and quantitative differences between strains. Differential abundance analysis led to detection of few quantitative differences between strains (Fig. 3), and qualitative differences also appear to be minor, with very few detectable on electrophoresis profiles. Venoms of the two strains therefore appeared to be very similar. As differential parasitism success depending on host can trigger rapid changes for some key venom factors (Cavigliasso et al., 2019), this similarity suggests that virulence differences between strains is caused by variation of few proteins. However, qualitative differences, even if rare, could be missed with the method used here as predicted digested proteome was made with consensus reference proteome. Therefore, some differences in protein sequences could not be revealed when identifying proteins. Proteins of very low abundance can also remain undetected.

We evidenced two types of changes between *C. typhae* strains. First, a large-scale change involving a high number of genes differentially expressed between the two strains, especially proviral genes at early parasitism stage. Second, specific changes on a few venom effectors. The scales are not mutually exclusive, as modification of few effectors can induce broader expression variations. Thus, they can both participate in the virulence-resistance interaction. Specific changes on few venom proteins have been reported for *L. boulardi* (Colinet et al., 2013), and if the same applies for *C. typhae* strain, CtV- virulence factors could be inefficient against SnR+ population. Similar to a gene-for-gene type of relation (Brading et al., 2002), CtV- strain would lack the specific virulence factor targeting SnR+ immune system, which would lead to failed parasitism. Indeed, virulence and resistance differences could generally be linked to a single locus (Carton et al., 2008). Alternatively, over-abundant proteins in CtV- venoms could act as elicitors of SnR+ immunity (Carton et al., 2008). If an over-abundant venom protein is detected by the host and triggers a high and efficient immune reaction, the interaction would become non-permissive (counter-adaptation called Effector Triggered Immunity, Remick et al. (2023)).

### 2.4. Molecular candidates to differential virulence

Host immune reaction can be modulated by venom proteins even in *Cotesia* species that inject polyDNAviruses (Pinto et al., 2021). One of the most known venom protein involved in immunosuppression is calreticulin, found in several species (Manzoor et al., 2016; Özbek et al., 2019) and able to suppress immune response of *P. xylostella* parasitized by *C. plutellae* (Cha et al., 2015). This protein was found in *C. typhae* venom, but was not differentially abundant between strains. As venom virulence could rely on the most abundant proteins (Tang et al., 2019), we decided to focus on the 5 most abundant proteins that displayed differences between strains. Among those, two proteins (KAG8039124.1 and KAG8039128.1) were over-abundant in CtV+ and identified as serine-protease inhibitors (serpins). Moreover, *in silico* verifications confirmed they exhibited the characteristic RCL (Reactive Center Loop) and global serpin-like structure (Colinet et al., 2009; Gettins, 2002), hinting toward a functional role in the host. Serpins were shown to be able to inhibit melanization processes (Asgari et al., 2003; Yokoi et al., 2017) and even in a case of CtV- failed parasitism, SnR+ capsules do not melanize compared to inert bead capsules (Gornard et al., 2024), indicating that melanization is disrupted in all cases. Consistent with other observations, melanization is not crucial for successful immunity (Colinet et al., 2009) but other serpins could still be involved in encapsulation *per se* (reviewed by Colinet et al., 2009; Danneels et al., 2010). Indeed, hemocyte spreading is altered by serpin activity of *Venturia canescens* (Beck et al., 2000) and *Pteromalus puparum* (Yan et al., 2023). Therefore, serpins are considered as serious candidate genes for *C. typhae* virulence difference, given that 2 of them are over-abundant in CtV+ venom.

Some venom proteins do not have a virulence role and rather participate in venom homeostasis, production, secretion and quality control (Moreau & Asgari, 2015; Poirié et al., 2014). Aminopeptidase (COTY_00001603), found in *Habrobracon* sp (Quicke & Butcher, 2021) and *C. chilonis* (Teng et al., 2017), was the most abundant *C. typhae* venom protein that differed in abundance between strains. It was 2.5 times more abundant than the second most abundant one, and was the fourth most abundant in the venom. Although this enzyme could be involved in host tissue degradation for offspring feeding or permeability to venom proteins (Scieuzo et al., 2021), it can also be linked to the processing of venom compounds, by cleaving them into biologically active molecules (Dani et al., 2003). Therefore, if CtV- venom lacks this enzyme while it is essential for its biological activity, it could be overall less efficient and not able to counteract SnR+ immune defenses.

Other venom proteins seem to play a role in passive defense, i.e. protecting the parasitoid eggs against encapsulation without disrupting the host cellular defenses. Immunoevasive proteins (IEP) of *Cotesia* venom are likely to act this way. In *C. kariyai*, they can protect the eggs from encapsulation by the host, but are inefficient when parasitizing a non-host (Hayakawa & Yazaki, 1997). This IEP is also found in *C. chilonis*, which resorts to passive evasion and immunosuppression, and could play the same role (Teng et al., 2017). Recent work showed that CtV- strain could use passive evasion strategies, consistent with the presence of an over-abundant IEP in its venom (KAG8035893.1) (Gornard et al., 2024, submitted). The passive evasion would then be efficient with SnR-, unable to recognize CtV- eggs as non-self, but not with SnR+, which could be able to detect them. In that case, the detection of CtV- by SnR+ host associated with weaker immunosuppressive virulence factors could lead to encapsulation of the eggs.

*C. chilonis* venom also contain metalloproteases, which have a nutrition role but can also participate in immunosuppression (Özbek et al., 2019), by modulating encapsulation via disruption of the Toll pathway (Scieuzo et al., 2021) or by blocking the initiation of defense mechanisms (Danneels et al., 2010). COTY_00009342 encodes a metalloprotease identified as putative venom that was not differentially abundant between strains, but represented an interesting candidate gene due to its over-expression in the CtV+ strain and its presence in a QTL. Proteomic differences between strains could reside in a qualitative rather than a quantitative aspect, possibly with a mutated, less efficient protein in CtV- strain.

Protein disulfide isomerases are enzymes capable of forming and rearranging disulfide bonds between cysteine amino acids of proteins. They are known in Conidae for allowing protein folding before secretion (Safavi-Hemami et al., 2016), and in parasitoid venom, they are thought to be involved in post-traductional processing of proteins (Liu et al., 2017). They have been identified in several Braconidae species (Becchimanzi et al., 2020; Teng et al., 2017), but without clear attributed function. Interestingly, that type of enzyme is not usually released in the extracellular lumen (Turano et al., 2002), but *C. typhae* venom disulfide isomerase (COTY_00016350) has an excretion signal, indicating its presence in the venom fluid is expected. Little is known about the role of such enzymes in parasitoid venoms, and it has to be studied precisely to be linked to virulence.

Finally, one candidate gene, COTY_00003145 was identified as a putative uncharacterized venom protein of *C. vestalis* but was not found in the venom gland proteome. This protein could still be injected in the host, either by the ovarian fluid or through teratocyte secretion, but its biological role also needs to be precisely identified.

No venom protein involved in provirus entry or expression in host cells was found to be differentially abundant between strains, consistent with the very low number of proviral genes differentially expressed between strains when focusing on the permissive host, SnR-, at both 24 and 96h.

This analysis focused on the most abundant venom proteins, but some factors can act at very low concentration (Poirié et al., 2014; Tang et al., 2019) and therefore constitute future research topics. Identification of novel virulence factors in venom could be achieved by combining proteomic analysis with venom gland transcriptome sequencing (Zhao et al., 2023).

# Conclusion

By combining different approaches and crossing them with genomic data, we revealed that virulence factor expression mostly happens during the nymphal stage, so that molecular weapons of the wasp can be ready upon emergence. We were also able to identify several candidate genes that could represent the molecular basis of virulence difference between two strains of *C. typhae*. Both proviral genes and venom proteins could be involved in the lower success of the CtV- strain parasitizing its French host. The two strains appear to rely on different virulence strategies, both with venom and proviral factors, and those strategies are more or less efficient, depending on the host strain. To a greater extent, this study provides a model for uncovering virulence factors in the genus *Cotesia*, and could lead to identifying the most efficient parasitoid strains against a given pest population, by using their venomic or proviral expression profiles.

## Supplementary information

Supplementary material 1: annotation of putative venom genes performed by iVenomDB (supmat1.xls)
Supplementary material 2: annotation of orthology terms performed by eggNOG-mapper (supmat2.xls)
Supplementary material 3: details about proteomic analysis and protein identification (supmat3.pdf)

## List of abbreviations:

CtV+: *Cotesia typhae*, highly virulent strain. CtV-: *Cotesia typhae*, less virulent strain. SnR+: *Sesamia nonagrioides*, resistant population. SnR-: *Sesamia nonagrioides*, permissive population. FSE: free-stage experiment. 24-PSE: parasitic stage experiment, 24 hours post-parasitism. 96-PSE: parasitic stage experiment, 96 hours post-parasitism. STAR: Spliced Transcripts Alignment to a Reference. BLAST: Basic Local Alignment Search Tool. AED: Annotation Edit Distance. GO: Gene Ontology. KEGG: Kyoto Encyclopedia of Genes and Genomes. iVenomDB: Insect Venom Database. VST: Variance Stabilizing Transformation. PBS: Phosphate Buffer Saline. SDS-PAGE: Sodium-Dodecyl-Sulfate PolyAcrylamide Gel Electrophoresis. LC-MS: Liquid Chromatography and Mass Spectrometry. DTT: Dithiothreitol. ANOVA: Analysis Of Variance. PCA: Principal Components Analysis. NCBI: National Center for Biotechnology. IGV: Integrative Genomics Viewer. IEP: ImmunoEvasive Protein.

## Declarations

### Acknowledgments

The authors thank Alain Peyhorgues for providing field-collected *S. nonagrioides* individuals for laboratory rearing at EGCE; Rémi Jeannette for *S. nonagrioides* rearing; Claire Capdevielle-Dulac and Sabrina Bothorel for *C. typhae* rearing at EGCE, Paul-André Calatayud and the Cotesia rearing team at ICIPE (Nairobi, Kenya) - Julius Obonyo, Josphat Akhobe, and Enock Mwangangi - for insect collection and shipment to refresh EGCE rearing; Jean-Luc Da Lage for SDS-PAGE analysis and advise. Work on Kenyan insects was done under the juridical framework of material transfer agreement CNRS 072057/IRD 302227/00.

### Author's contributions

F.M. and L.K. conceived and planned the project. S.G., P.V. and F.L. collected and analyzed transcriptomic data. S.G. collected venom glands. T.B. and S.G. analyzed proteomic data. F.M. and L.K. overviewed and contributed to data analysis. S.G., F.M. and L.K. wrote the manuscript, with the contribution of T.B. on Methods section. All authors approved the final manuscript.

### Funding

This study was co-funded by the French National Research Agency (ANR) and the National Agency for Biodiversity (AFB) (grant CoteBio ANR17-CE32-0015-02 to L.K.), and by the Ecole doctorale 227 MNHN-UPMC Sciences de la Nature et de l'Homme: évolution et écologie.

### Availability of data and materials

The RNA-sequencing data for this study have been deposited in the European Nucleotide Archive (ENA) at EMBL-EBI under accession number PRJEB75362 and PRJEB75367 (release date: 31/12/2024).
The mass spectrometry proteomics data have been deposited to the ProteomeXchange Consortium via the PRIDE (Perez-Riverol et al., 2022) partner repository with the dataset identifier PXD051968.

### Ethics approval and consent to participate

Not applicable. No human tissue or data was used in the experiments, and the animal tissues were used in compliance with institutional, national and international ethical guidelines.

### Consent for publication

Not applicable.

### Competing interests

The authors declare no competing interests.

### Author details

[1] EGCE, Université Paris-Saclay, CNRS, IRD, UMR Évolution, Génomes, Comportement et Écologie, 91190 Gif-sur-Yvette, France

[2] PAPPSO, Université Paris-Saclay, INRAE, CNRS, AgroParisTech, GQE - Le Moulon, 91190, Gif-sur-Yvette, France